\documentclass[useAMS,usenatbib]{mn2e}
\usepackage{psfig,amssymb,amsmath,rotating,nicefrac,longtable,color,natbib,setspace,threeparttable}
\def\deg{$^{\circ}$}
\def\marcmin{^{\prime}} 
\def\marcsec{^{\prime\prime}}

\def\psq{$^{-2}$}
\def\zth{$z\approx3$}
\def\lone{\textsc{lbg\_pri{\footnotesize1}}}
\def\ltwo{\textsc{lbg\_pri{\footnotesize2}}}
\def\lthree{\textsc{lbg\_pri{\footnotesize3}}}
\def\ldrop{\textsc{lbg\_drop}}
\def\lya{Ly$\alpha$}

\def\overh{h^{-1}}
\def\ps{s$^{-1}$}
\def\hmpc{~\overh\mathrm{Mpc}}
\def\kps{\mathrm{kms}^{-1}}
\def\arcmin{$^{\prime}$}
\def\arcsec{$^{\prime\prime}$}

\def\vdisp{\sqrt{\left<w_z^2\right>}}

\def\totlbg{2,135}
\def\numlbg{1,994}

\def\meanz{2.79}
\def\vwprn{3.46}
\def\vwprne{0.41}
\def\vwpgam{1.52}
\def\vwpgame{0.13}
\def\vwpic{5.33}
\def\vxiic{0.024}
\def\vbias{2.37}
\def\vbiase{0.21}
\def\vhmass{11.57}
\def\vhmasse{0.15}
\def\vhminm{11.13}
\def\vhminme{0.18}
\def\vhocc{1.2}
\def\vhocce{0.6}
\def\kwprn{3.98}
\def\kwprne{0.32}
\def\kwpgam{1.58}
\def\kwpgame{0.13}
\def\kwpic{7.18}
\def\kxiic{0.064}
\def\kbias{2.78}
\def\kbiase{0.13}
\def\khmass{11.69}
\def\khmasse{0.10}
\def\khminm{11.30}
\def\khminme{0.10}
\def\khocc{1.8}
\def\khocce{0.6}
\def\cwprn{3.83}
\def\cwprne{0.24}
\def\cwpgam{1.60}
\def\cwpgame{0.09}

\def\cbias{2.59}
\def\cbiase{0.13}
\def\chmass{11.73}
\def\chmasse{0.07}
\def\chminm{11.33}
\def\chminme{0.09}
\def\chocc{2.0}
\def\chocce{0.5}
\def\vdispar{420}
\def\vdisparup{+140}
\def\vdispardown{-160}
\def\infpar{0.38}
\def\infpare{0.19}
\def\infparoma{0.38}
\def\infparomaup{+0.15}
\def\infparomadown{-0.13}
\def\infparomb{0.38}
\def\infparombup{+0.16}
\def\infparombdown{-0.09}
\def\ompara{0.08}
\def\omparaup{+0.22}
\def\omparadown{-0.08}
\def\omparb{0.30}
\def\omparbup{+0.32}
\def\omparbdown{-0.18}
\def\fpar{0.99}
\def\fpare{0.50}
\def\fsig{0.26}
\def\fsige{0.13}

\defcitealias{2011MNRAS.414....2B}{Paper I}
\defcitealias{2011MNRAS.414...28C}{Paper II}

\title[VLT LBG Redshift Survey III]{The VLT LBG Redshift Survey - III. The clustering and dynamics of Lyman-break galaxies at $z\sim3$
\thanks{Based on data obtained with the NOAO Mayall 4m Telescope at Kitt Peak National Observatory, USA (programme ID: 06A-0133), the NOAO Blanco 4m Telescope at Cerro Tololo Inter-American Observatory, Chile (programme IDs: 03B-0162, 04B-0022) and the ESO VLT, Chile (programme IDs: 075.A-0683, 077.A-0612, 079.A-0442).}}
\author[R. Bielby et al.]{R. Bielby$^{1}$\thanks{E-mail:
rmbielby@gmail.com (RMB)}, M. D. Hill$^{1}$, T. Shanks$^{1}$, N. H. M. Crighton$^{1,2}$, L. Infante$^{3}$,
\newauthor C. G. Bornancini$^4$, H. Francke$^3$, P. H\'{e}raudeau$^{5,6}$, D. G. Lambas$^4$, N. Metcalfe$^1$
\newauthor D. Minniti$^{3,7,8}$, N. Padilla$^3$, T. Theuns$^{9,10}$, P. Tummuangpak$^{1}$, P. Weilbacher$^{11}$\\
$^{1}$Durham University, South Road, Durham, DH1 3LE, United Kingdom\\
$^{2}$Max Planck Institute for Astronomy, Kšnigstuhl 17, D-69117 Heidelberg, Germany\\
$^{3}$Departamento de Astronom\'\i a y Astrof\'isica, Pontificia Universidad Catolica de Chile, Casilla 306, Santiago 22, Chile\\
$^{4}$Instituto de Astronom\'ia Te\'orica y Experimental (CONICET-UNC), Observatorio Astron\'omico de C\'ordoba, Laprida 854, \\X5000BGR, C\'ordoba, Argentina\\
$^{5}$Argelander Institut f\"ur Astronomie, Auf dem H\"ugel 71, 53121 Bonn, Germany\\
$^{6}$Kapteyn Astronomical Institute, University of Groningen, PO Box 800, 9700 AV, Groningen, The Netherlands\\
$^{7}$Vatican Observatory, V00120 Vatican City State, Italy\\
$^{8}$Department of Astrophysical Sciences, Princeton University, Princeton NJ 08544-1001, USA\\
$^9$Institute for Computational Cosmology, Department of Physics, University of Durham, South Road, Durham, DH1 3LE, UK\\
$^{10}$Department of Physics, University of Antwerp, Campus Groenenborger, Groenenborgerlaan 171, B-2020 Antwerp,  Belgium\\
$^{11}$Leibniz-Institut f\"{u}r Astrophysik Potsdam (AIP), Germany
}
\begin{document}

\date{First draft 2012 April 5}

\pagerange{\pageref{firstpage}--\pageref{lastpage}} \pubyear{2012}

\maketitle

\label{firstpage}

\begin{abstract}
We present a catalogue of \totlbg\ galaxy redshifts from the VLT LBG Redshift Survey (VLRS), a spectroscopic survey of $z\approx3$ galaxies in wide fields centred on background QSOs. We have used deep optical imaging to select galaxies via the Lyman break technique. Spectroscopy of the Lyman-break Galaxies (LBGs) was then made using the VLT VIMOS instrument, giving a mean redshift of $z=\meanz$. We analyse the clustering properties of the VLRS sample and also of the VLRS sample combined with the smaller area Keck based survey of Steidel et al. From  the semi-projected correlation function, $w_p(\sigma)$, for the VLRS and combined surveys, we find that the results are well fit with a single power law model, with clustering scale lengths of $r_0=\vwprn\pm0.41$ and $\cwprn\pm0.24~h^{-1}{\rm Mpc}$ respectively. We note that the corresponding  combined $\xi(r)$ slope is flatter than for local galaxies at $\gamma=1.5 - 1.6$ rather than $\gamma=1.8$. This flat slope is confirmed by the $z$-space correlation function, $\xi(s)$, and in the range $10<s<100\hmpc$ the VLRS shows an $\approx2.5\sigma$ excess over the $\Lambda$CDM linear prediction. This excess may be consistent with recent evidence for non-Gaussianity in clustering results at $z\approx1$. We then analyse the LBG $z$-space distortions using the 2-D correlation function, $\xi(\sigma,\pi)$, finding for the combined sample a large scale infall parameter of $\beta=\infpar\pm\infpare$ and a velocity dispersion of $\vdisp=\vdispar^{\vdisparup}_{\vdispardown}\kps$. Based on our measured $\beta$, we are able to determine the gravitational growth rate, finding a value of $f(z=3)=\fpar\pm\fpare$ (or $f\sigma_8=\fsig\pm\fsige$), which is the highest redshift measurement of the growth rate via galaxy clustering and is consistent with $\Lambda$CDM. Finally, we constrain the mean halo mass for the LBG population, finding that the VLRS and combined sample suggest mean halo masses of log$(M_{DM}/M_\odot)=\vhmass\pm\vhmasse$ and log$(M_{DM}/M_\odot)=\chmass\pm\chmasse$ respectively.
\end{abstract}


\section{Introduction}

The large scale structure of matter presents a crucial guide in understanding the nature and evolution of the Universe. In $\Lambda$CDM, structure in the Universe grows hierarchically through gravitational instability (e.g. \citealt{mowhite96,1998ApJ...499...20J,2006Natur.440.1137S}) and testing this model requires the measurement of the matter clustering and the growth of structure across cosmic time (e.g. \citealt{2005Natur.435..629S,2008MNRAS.391.1589O,2009MNRAS.400.1527K}). We are limited however in our ability to trace the structure of mass given that observations suggest that $\approx75\%$ of the mass density of the Universe is in the form of dark matter.


Although large photometric surveys are beginning to map the overall matter density distribution via its lensing signature \citep[e.g.][]{2007Natur.445..286M,2012MNRAS.tmp.2386H}, at present the primary tool in the statistical analysis of the distribution of matter in the Universe remains the study of the clustering statistics of selected galaxy populations. A given galaxy population traces the peaks in the matter distribution and hence provides a biased view of the matter density, which nevertheless can be used to follow the overall growth of structure. 

At low-redshift, magnitude limited galaxy samples have provided significant tools in probing the clustering properties of the galaxy population \citep[e.g.][]{2002MNRAS.332..827N,hawkins03}, whilst at higher redshifts, photometric selections are required to isolate the required redshift range, for example the Luminous Red Galaxy (LRG), $BzK$, Extremeley Red Object (ERO), Distant Red Galaxy (DRG) selections. At $z>2$, identifying galaxy populations is primarily reliant on the Lyman-break galaxy \citep[LBGs; e.g. ][]{1996MNRAS.283.1388M,steidel96,steidel99} and the Ly-$\alpha$ emitter \citep[LAEs; e.g.][]{1998AJ....115.1319C,2006ApJ...642L..13G,2007ApJ...671..278G,2008ApJS..176..301O} selections.

In particular, the Lyman-break technique has proven highly successful in surveying the $z>2$ Universe. The LBGs represent a large population of star-forming galaxies in the high redshift Universe. In comparison to LAEs, the LBG selections offer the advantage of both a contiguous and broader range of redshifts, whilst the typically brighter apparent magnitudes of LBGs mean that it is possible to obtain much more detailed information on stellar populations for individual objects, and also to measure a range of interstellar absorption features in rest-frame UV spectra.


\citet{steidel03} presented a large survey of LBGs in the redshift range $2.5<z<3.5$, identifying $\approx800$ such galaxies based on spectroscopic observations using the Keck {\sc I} telescope. \citet{adelberger03} used this sample to measure the auto-correlation of the LBGs for comparison to the cross-correlation between LBGs and gas as traced by the {\sc HI} and {\sc CIV} absorption features in quasar sightlines. They fit the auto-correlation function with a simple power-law and reported a clustering length for the galaxies of $r_0=3.96\hmpc$ (with a slope of $\gamma=1.55$).

\citet{2005ApJ...619..697A} continued from the previous work, presenting an analysis of the clustering properties of galaxies selected photometrically with three different methods including the LBG method. Based on both photometric and spectroscopic samples they found a clustering length of $r_0=4.0\hmpc$ and slope of $\gamma=1.6$, consistent with the previous Keck analyses. Comparison to numerical simulations suggested that such clustering properties were consistent with the LBGs residing in dark matter (DM) halos with average masses of $10^{11.2-11.8}\mbox{M}_\odot$, concluding that the typical LBG will have evolved into an elliptical galaxy at $z=0$ and will have an early type stellar population by $z\sim1$. This was however contradicted by \citet{2008ApJ...679.1192C} and \citet[][Paper I]{2011MNRAS.414....2B}, both of whom showed that the clustering evolution of the LBG population may be more complicated, but is likely to produce typical $L^\star$ galaxies at $z\sim0$. Interestingly, this is well complemented by the findings of \citet{2007ApJ...654..138Q,2008ApJ...685L...1Q} who show that optically-faint/$K$-band bright galaxies at $z\sim2-3.5$ are far more highly clustered than the optically bright LBG population, and hence suggest that it is this optically faint population missed by the LBG selection that evolves into the massive elliptical population at $z\sim0$. Other observations show consistent measurements of the halo masses in which LBGs reside \citep[e.g.][]{foucaud03,2009A&A...498..725H,2011ApJ...737...92S,2012ApJ...752...39T,2012arXiv1208.2097J}. Similarly, the complexity of the evolutionary track of LBGs is supported by recent simulations. For example, \citet{2012MNRAS.423.3709G} find that LBGs can be successfully simulated as starbursts triggered by minor mergers, with host halo masses of $\sim3\times10^{11} h^{-1}M_{\odot}$. These are marginally preferentially disk dominated systems at $z\sim3$ that evolve into Milky Way mass galaxies with 50:50 bulge-disk dominated systems.


Taking the galaxy clustering measurements, it is possible to measure the large scale dynamics of the galaxy population through redshift space distortions. For instance, \citet{daangela05b} took the \citet{steidel03} Keck spectroscopic sample and used the clustering properties of the LBG population to constrain the cosmological density parameter, $\Omega_m$, and the bulk motion properties of the large scale structure at $z\approx3$. By measuring the 2D clustering of the galaxy distribution, they placed constraints on the infall parameter of $\beta(z=3)=0.25^{+0.05}_{-0.06}$ and on the mass density of $\Omega_m(z=0)=0.55^{+0.45}_{-0.16}$. However, the small fields of view available from the \citet{steidel03} survey meant the authors could not solve for both the bulk motion and the velocity dispersion, which are degenerate, severely limiting the scope of the results. \citetalias{2011MNRAS.414....2B} improved on these results by combining the data with first galaxy sample from the VLT LBG Redshift Survey (VLRS). By adding $\approx$ 1,000 galaxies to the $z\approx3$ sample of \citet{steidel03} data across much larger fields, they measured the clustering and dynamics of the $z\approx3$ LBG population. With the wider fields available, \citetalias{2011MNRAS.414....2B} were able to begin to probe both the small scale peculiar velocity field and the large scale bulk motion field. The authors showed that the redshift space distortions of the $z\sim3$ galaxy population are well fit by a model with an infall parameter of $\beta=0.48\pm0.17$, which they went on to show is consistent with the standard $\Lambda$CDM cosmology. This was similar to a number of other works performed based on redshift distortions at lower-redshifts, for example \citet{2006PhRvD..74l3507T, ross07, 2008Natur.451..541G, 2009JCAP...10..004S}, where contraints have been placed on the growth of structure. However, few constraints on this important cosmological measure are available at redshift of $z\gtrsim1$.

In this paper, we add to the previous results of the VLT LBG Redshift Survey (VLRS) presented in \citetalias{2011MNRAS.414....2B}, \citet[][Paper II]{2011MNRAS.414...28C} and \citet{2011Msngr.143...42S}. We present new spectroscopic LBG data obtained using the VLT VIMOS instrument, more than doubling both the area covered and the number of spectroscopically confirmed galaxies in the survey. We use the updated survey to measure the clustering and dynamical properties of the \zth\ LBG population. Throughout this paper, we use a cosmology given by $H_0=70\kps$, $\Omega_m=0.3$, $\Omega_\Lambda=0.7$ and $\sigma_8=0.8$. In addition distances are quoted in comoving coordinates in units of $\hmpc$ unless otherwise stated.

\section{Observations}

\subsection{Survey overview}

\begin{table*}
\centering
\caption[A summary of the fields making up our $z\approx3$ LBG survey.]{A summary of the fields making up our \zth\ LBG survey. The table gives the name, coordinates and redshift of the QSO on which the fields are roughly centred, as well as the number of subfields (individual VLT VIMOS pointings) with spectroscopic data. The first block of fields were presented by \citetalias{2011MNRAS.414....2B}, the second block are presented in this paper.}
\label{t-qsofields}
\begin{tabular}{lcccccc}
\\
\hline
Field & RA$^a$ & Dec$^a$ & $z$ $^b$ & Subfields & Reference\\
\hline
Q0042--2627  & 00:44:33.9 & -26:11:21 & 3.29 & 4 & \citetalias{2011MNRAS.414....2B}\\
J0124+0044   & 01:24:03.8 & +00:44:33 & 3.84 & 4 & \citetalias{2011MNRAS.414....2B}\\
HE0940--1050 & 09:42:53.4 & -11:04:25 & 3.05 & 3 & \citetalias{2011MNRAS.414....2B}\\
J1201+0116   & 12:01:44.4 & +01:16:12 & 3.23 & 4 & \citetalias{2011MNRAS.414....2B}\\
PKS2126--158 & 21:29:12.2 & -15:38:41 & 3.28 & 4 & \citetalias{2011MNRAS.414....2B}\\
\hline
 & & & & 19  & \\
\hline
Q2359+0653   & 00:01:40.6 & +07:09:54 & 3.23 & 4 & This work \\
Q0301--0035  & 03:03:41.0 & -00:23:22 & 3.23 & 4 & This work \\
Q2231+0015   & 22:34:09.0 & +00:00:02 & 3.02 & 3 & This work \\
HE0940--1050 & 09:42:53.4 & -11:04:25 & 3.05 & 6 & This work \\
Q2348-011    & 23:50:57.9 & -00:52:10 & 3.02 & 9 & This work \\
\hline
 & & & & 26 & \\\hline
\multicolumn{7}{l}{$^a$ J2000 coordinates of QSO; not necessarily the exact centre of the observed field.}\\
\multicolumn{7}{l}{$^b$ redshift of the central quasar}\\
\\
\end{tabular}
\end{table*} 

In order to facilitate an investigation of how \zth\ galaxies interact with gas in the intergalactic medium (IGM), the survey comprises observations of several target fields centred on bright $z>3$ quasars, since features in the QSO spectra can provide information on the local IGM. \citetalias{2011MNRAS.414....2B} presented the first 5 fields of the survey, centred on the following quasars: Q0042--2627 ($z=3.29$), J0124+0044 ($z=3.84$), HE0940--1050 ($z=3.05$), J1201+0116 ($z=3.23$) and PKS2126--158 ($z=3.28$), hereafter referred to by only the right-ascension component of these names. A spectroscopic survey of each of these quasar fields was carried out with the Visible Multi-Object Spectrograph (VIMOS) on the European Southern Observatory's Very Large Telescope (VLT) in Chile (during the ESO periods 75-79). Each field consisted of four sub-fields (individual pointings with the VLT spectrograph), except for HE0940 where only three sub-fields were available at the time of their publication. A VIMOS pointing has a field of view of $16\marcmin \times 18\marcmin$ (see \S\ref{ss-specobs}), therefore each quasar field covered $\approx32\marcmin \times 36\marcmin$, or $\approx0.32$ deg$^2$, except for HE0940 which with 3 sub-fields covered $\approx0.24$ deg$^2$.

Building on this initial dataset, we present the continuation of these
observations since incorporating ESO periods 81 and 82. We have added a
further 6 sub-fields to HE0940, tripling its previous area, as well as
observations of 4 new fields, around the quasars Q2359+0653 ($z=3.23$),
Q0301--0035 ($z=3.23$), Q2231+0015 ($z=3.02$) and Q2348-011 ($z=3.02$),
with 4, 4, 3 and 9 sub-fields respectively. Table \ref{t-qsofields}
summarises all the fields of the survey. This
includes those presented by \citetalias{2011MNRAS.414....2B}, covering
1.52 deg$^2$, and those presented here, which take the total observed
area to 3.6 deg$^2$, more than doubling the previous size.

\subsection{Imaging}

\subsubsection{Observations and data reduction}

The selection of \zth\ LBG candidates was performed using photometry from optical broadband imaging. The imaging data for Q2359 and Q0301 were acquired with the Mosaic wide-field imager on the 4m Mayall telescope at Kitt Peak National Observatory (KPNO) in September 2005. The Q2231 data are from the Wide Field Camera on the 2.5m Isaac Newton Telescope (INT) on La Palma, and were observed in August 2005. All of these observations were carried out in the $U$, $B$ and $R$ bands. 

The MOSAIC imager at KPNO consists of 8 2k$\times$4k CCDs arranged into an 8k$\times$8k square. With a plate scale of 0.26\arcsec/pixel, this gives a field of view of 36\arcmin$\times$36\arcmin. There are 0.5--0.7mm gaps between the chips, corresponding to gaps of 9--13\arcsec\ on-sky, so a dithering pattern was used during the observations to provide complete field coverage. $U$, Harris $B$ and Harris $R$ filters were used. 

The MOSAIC data were reduced using the \texttt{mscred} package in \textsc{iraf}. The reduction process is described by \citetalias{2011MNRAS.414....2B}, however we briefly outline the procedure here. Initially a master bias frame is produced for each night's observing. The dome flats and sky flats were then processed using the \texttt{ccdproc} and \texttt{mcspupil} routines, subtracting the bias and eliminating the faint 2600-pixel pupil image artefact. The object frames were processed similarly, subtracting the bias and pupil image, and then were flat-fielded using the dome and sky flats. Bad pixels and cosmic rays were masked out of the science frames using the \texttt{crreject}, \texttt{crplusbpmask} and \texttt{fixpix} procedures. Finally, the \textsc{swarp} software package \citep{swarp} was used to resample and co-add the frames, producing a final science image.

The HE0940 and Q2348 data were acquired with the MegaCam imager on the 3.6m Canada-France-Hawai'i Telescope (CFHT). HE0940 was observed using the CFHT $u^*$, $g'$, $r'$, $i'$ and $z'$ bands in April 2004 as part of the observing run 2004AF02 (PI: P. Petitjean), whilst Q2348 was observed in the $u^*$, $g'$, $r'$ and $z'$ bands over the period August-December 2004 as part of the observing run 2004BF03 (PI: P. Petitjean). Table \ref{t-imobs} gives full details of all the imaging data. For this work we used pre-reduced individual exposures provided by the Elixir system at the CFHT Science Archive, which we then stacked using the \textsc{scamp} \citep{scamp} and \textsc{swarp} \citep{swarp} software packages.

\begin{table*}
\centering
\caption{Details of imaging observations for the LBG target fields presented in this paper.}
\label{t-imobs}
\begin{tabular}{lccllllll}
\\
\hline
Field   & RA          & Dec         & Instrument       &Band&Exposure& Seeing     &Completeness & Dates\\
        &\multicolumn{2}{c}{(J2000)}&                  &    & (ks)   &            &(50\% Ext/PS)& \\
\hline
Q2359   & 00:01:44.85 & +07:11:56.0 & Mosaic (KPNO)    & $U$ & 19.2 &  1.46\arcsec\ & 24.76/25.18 & 29--30 Sep 2005\\
        &             &             &                  & $B$ & 7.2  &  1.45\arcsec\ & 25.28/25.73 & \\
        &             &             &                  & $R$ & 6.0  &  1.15\arcsec\ & 24.74/25.20 & \\
\hline
Q0301   & 03:03:45.27 &  -00:21:34.2 & Mosaic (KPNO)   & $U$ & 19.2 &  1.34\arcsec\ & 24.93/25.34 & 29--30 Sep 2005\\
        &             &              &                 & $B$ & 6.4  &  1.28\arcsec\ & 25.51/26.04 &\\
        &             &              &                 & $R$ & 4.8  &  1.19\arcsec\ & 24.59/25.17 &\\
\hline
Q2231   & 22:34:28.00 & +00:00:02.0  & WFCam (INT)     & $U$ & 54.0 &  1.23\arcsec\ & 25.08/25.52 & 30 Aug 2005\\
        &             &              &                 & $B$ & 13.2 &  1.01\arcsec\ & 25.88/26.12 &  \\
        &             &              &                 & $R$ & 19.2 &  1.01\arcsec\ & 24.75/25.24 & \\
\hline
HE0940  & 09:42:53.06 & -11:02:56.9  & MegaCam (CFHT)  & $u^*$ & 6.8  &  0.99\arcsec\ & 25.39/25.93 & 14, 21--27 Apr 2004\\
        &             &              &                 & $g'$ & 3.1  &  0.86\arcsec\ & 25.54/26.05 & \\
        &             &              &                 & $r'$ & 3.7  &  0.85\arcsec\ & 25.08/25.65 & \\
\hline
Q2348   & 23:50:57.90 & -00:52:09.9  & MegaCam (CFHT)  & $u^*$ & 9.9  &  0.78\arcsec\ & 25.97/26.62 & 19--20 Aug, \\
        &             &              &                 & $g'$ & 5.5  &  0.79\arcsec\ & 25.71/26.29 & 7--10 Nov, \\
        &             &              &                 & $r'$ & 4.4  &  0.75\arcsec\ & 25.22/25.80 & 15 Dec 2004\\
\hline
\end{tabular}
\end{table*} 

The Wide Field Camera (WFCam) on the INT comprises 4 2k$\times$4k CCDs. These are arranged into a 6k$\times$6k block with a 2k$\times$2k square missing. With $\approx1\marcmin$ gaps between chips and a pixel scale of 0.33\arcsec/pixel, WFCam has a total FoV of $\approx$ 34\arcmin$\times$34\arcmin\ (0.32 deg$^2$); however, accounting for the incomplete coverage of the field, the total observing area is reduced to 0.28 deg$^2$.

The WFCam observations of Q2231 were made using the RGO $U$, Harris $B$ and Harris $R$ filters. The RGO $U$ filter has a central wavelength of 3581\AA\ and a FWHM of 638\AA, making it very similar to the $U$ band filter used at KPNO (centre 3552\AA, FWHM 631\AA). The $B$ and $R$ band filters were the same as at KPNO. Therefore, given that the filters are so similar, we will use the same $UBR$ selection criteria when identifying LBG candidates in either the Mosaic or WFCam datasets. 

Initial data reduction, including bias removal, flat-fielding and photometric calibration, was performed by the Cambridge Astronomical Survey Unit (CASU). Astrometry calibration and exposure stacking was performed using the \textsc{scamp} and \textsc{swarp} packages.

\subsubsection{Filters}

\begin{figure}
\centering
\includegraphics[width=0.45\textwidth]{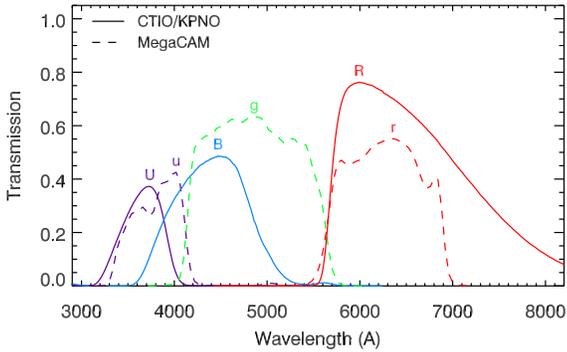}
\caption[]{The transmission profiles for the filter combinations used at KPNO/CTIO (solid curves - $UBR$) and CFHT (dashed curves - $ugr$).}
\label{fig:filters}
\end{figure}

As described above, our observations incorporate two different filter combinations. We show both the MegaCam $u^*g'r'$ and CTIO/KPNO $UBR$ filter profiles in Fig.~\ref{fig:filters}. The MegaCAM filters have central wavelengths of 3740\AA\, 4870\AA\ and 6250\AA\ for the $u^*$, $g'$ and $r'$ filters respectively, whilst the CTIO/KPNO filters have central wavelengths of 3570\AA\, 4360\AA\ and 6440\AA\ for the $U$, $B$ and $R$ filters respectively. These are both well suited to isolating the Lyman break in $z\sim3$ galaxies, however the MegaCAM $u^*$ and $g'$ filters are marginally redder than the Johnson-Cousins $U$ and $B$ filters.

Conversions from the MegaCAM filter set to the SDSS filter set are given in the CFHT MegaCAM technical documentation, whilst conversions from the SDSS filter set to the Johnson-Cousins system are given by \citet{1996AJ....111.1748F}. Combining these two sets of relations gives the following conversions between the two filter systems used in this work:

\begin{equation}
(u^*-g') = 1.05(U-B) + 1.10
\end{equation}

\begin{equation}
(g'-r') = 0.57(B-R) - 0.22
\end{equation}

These relations are used throughout this paper where comparing the MegaCAM and Johnson-Cousins colours.

\subsubsection{Photometry}

Photometric zeropoints for the imaging fields were determined from standard star observations carried out as part of each of the imaging runs. The standard star fields were reduced in the same way as the science frames to ensure consistency. Source detection in the science images was performed with SExtractor (Bertin \& Arnouts, 1996), using a 1.5$\sigma$ detection threshold and a 5-pixel minimum size. 

\begin{figure*}
\centering
\includegraphics[width=\textwidth]{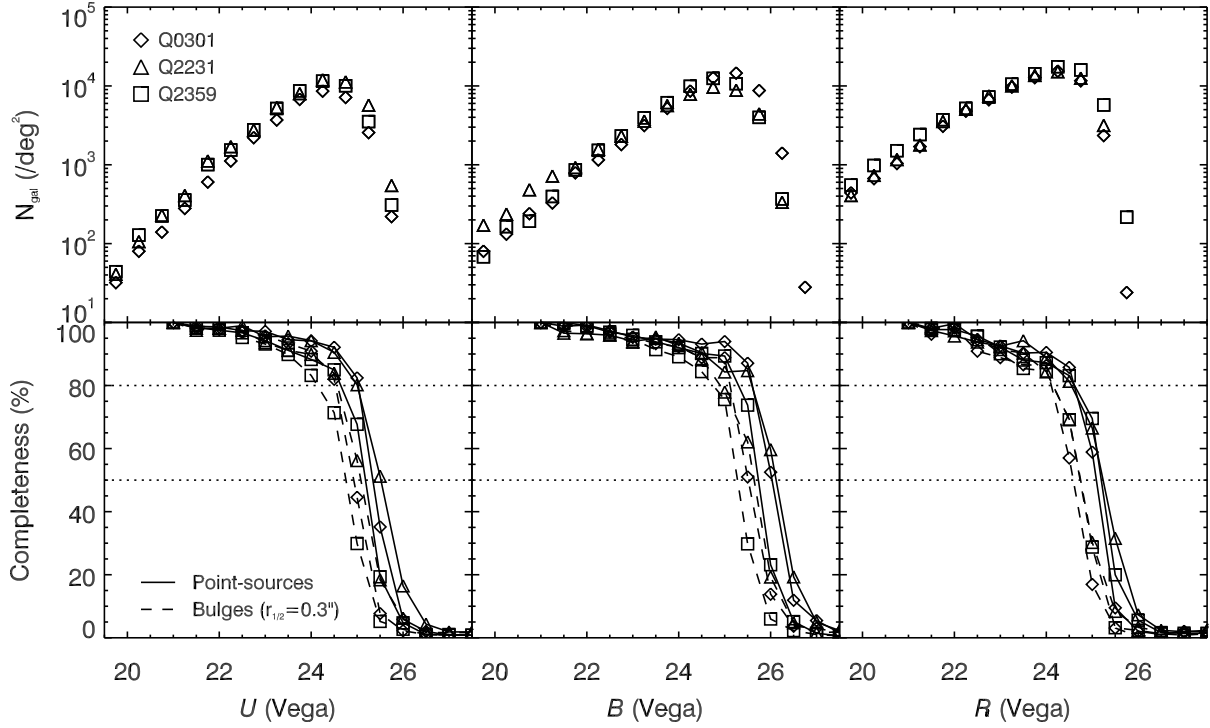}
\caption[Number counts and completeness measurements for LBG fields I]{\emph{Upper panel:} Galaxy number counts in the $U$ (left), $B$ (middle) and $R$ (right) band imaging for the fields Q0301 (diamonds), Q2231 (triangles) and Q2359 (squares). \emph{Lower panel:} Estimated completeness for each of the above bands based on simulated point sources (solid lines) and Vaucouleurs profile sources (dashed lines).}
\label{fig:vega_nc_cmp}
\end{figure*}

The $U_{Vega}$, $B_{Vega}$ and $R_{Vega}$ band galaxy number counts in the Q0301 (diamonds), Q2231 (triangles) and Q2359 (squares) LBG fields are shown in the top panels of Fig.~\ref{fig:vega_nc_cmp}. Stars were removed from these counts at magnitudes brighter than $\approx22$ using a limit on the measured half-light radius of the sources. At fainter magnitudes, no attempt to remove stars from the counts was made, as the smallest extended sources become unresolved at the PSF of our fields at such magnitudes. We also show completeness estimates for each image in each field. These are estimated by placing simulated sources at random positions in a given image and measuring the fraction that are successfully extracted with SExtractor (using the same extraction parameters as used to create the full catalogues). In each case we estimate the completeness using both simulated point-sources and extended sources, where the extended sources are modelled by a de Vaucouleurs $r^{1/4}$ profile with a half-light radius of $r_{1/2}=0.3''$. In both cases, the simulated source is convolved with the image PSF before being added to the observation.

The results of the completeness estimates for the $UBR_{Vega}$ filter fields are shown in the lower panels of Fig.~\ref{fig:vega_nc_cmp}. The same symbols as the top panels are used for the different fields, whilst the dashed curves show the completeness estimates based on the extended sources and the solid curves show the completeness for the simulated point-sources. The 50\% limits completion estimates (equivalent to $\approx3\sigma$ detection limits) are given in Table~\ref{t-imobs}. Comparing the completeness measurements across the fields, the measurements are relatively consistent with the imaging in each field reaching comparable depths. We note that given the compact nature of the LBG targets, the point source completeness levels should be a good representation of the true completeness. As such all our fields reach depths of $R>25$.

\begin{figure*}
\centering
\includegraphics[width=\textwidth]{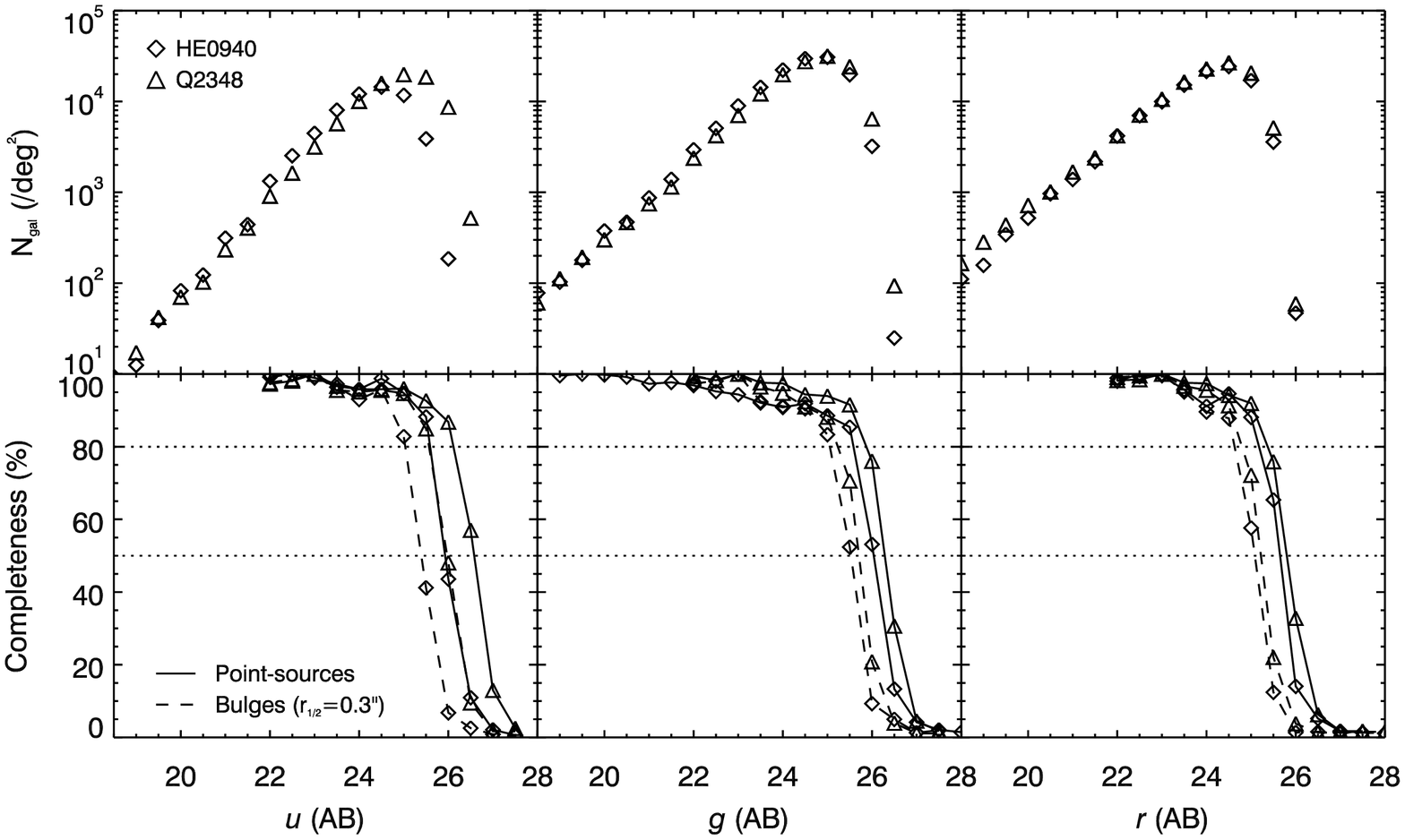}
\caption[$B$ band number counts for the LBG fields]{Number counts (top) and completeness estimates (bottom) in the $u$ (left), $g$ (middle) and $r$ (right) bands from the MegaCAM imaging on the HE0940-1050 and Q2348-011 fields.}
\label{fig:megacam_nc_cmp}
\end{figure*}
 
We show the galaxy number counts (top panels) and completeness estimates (lower panels) for the MegaCAM fields in Fig.~\ref{fig:megacam_nc_cmp}. Again the symbols are consistent between top and lower panels with the diamonds showing the results for the HE0940 field and the triangles showing the Q2348 field. As before, the solid lines in the lower panels show the completeness estimates for the point-like sources and the dashed lines show the same for the extended sources (which use the same de Vaucouleurs profile as used for the $UBR_{Vega}$ fields). Comparing the two fields to each other, the depths reached are comparable in each band, although the HE0940 is marginally less deep in the $u$ band by $\approx0.5$ mag.

\subsection{Candidate selection} \label{s-candsel}

\subsubsection{$UBR$ selection}

In the Q2359, Q0301 and Q2231 fields, we selected LBG candidates based on their $U$, $B$ and $R$ photometry. The criteria used were the same as those used by \citetalias{2011MNRAS.414....2B}, which are based on those of \citet{steidel03}. There are 4 groups to the selection, designated \lone, \ltwo, \lthree\ and \ldrop\ and defined as follows:

\begin{description}

\item[\lone] \hfill
\begin{itemize} 
\item $23<R\leq25$
\item $0.5<(U-B)<4.0$
\item $(B-R)<0.8(U-B)+0.6 $
\item $(B-R)<2.2$
\end{itemize}

\item[\ltwo] \hfill
\begin{itemize}
\item $23<R\leq25$
\item $(U-B)>0.0$
\item $(B-R)<0.8(U-B)+0.8 $
\item $-1<(B-R)<2.7 $
\item $\notin$ \lone
\end{itemize}

\item[\lthree] \hfill
\begin{itemize}
\item $23<R\leq25$
\item $-0.5<(U-B)<0.0 $
\item $-1.0<(B-R)<0.8(U-B)+0.6 $
\item $\notin$ \{\lone,\ltwo\}
\end{itemize}

\item[\ldrop] \hfill
\begin{itemize}
\item $23<R\leq25$
\item $0.5<(B-R)<2.2 $
\item No detection in $U$
\end{itemize}
\end{description}

The first 3 groups represent an order of priority --- that is, \lone\ candidates are considered more likely to be \zth\ LBGs than e.g.\ \lthree\ candidates. This is because whereas \lone\ tends to select outliers in the $UBR$ colour--colour plot, the lower priority groups select objects increasingly close to the colour region populated by stars and lower redshift galaxies, and therefore suffer from increased contamination from lower-redshift interlopers. 

The fourth group is somewhat separate, being for galaxies which are not detected in the $U$ band. Such sources may be excellent LBG candidates, since it may be that the presence of the Lyman limit in the $U$ band has made the galaxy extremely faint in this band, such that it `drops out' below the magnitude limit. However, the \ldrop\ population is also likely to suffer from contamination, in this case because objects with no counterpart in 1 of the 3 bands have a higher chance of being spurious sources.

\begin{figure}
\centering
\includegraphics[width=8.0cm]{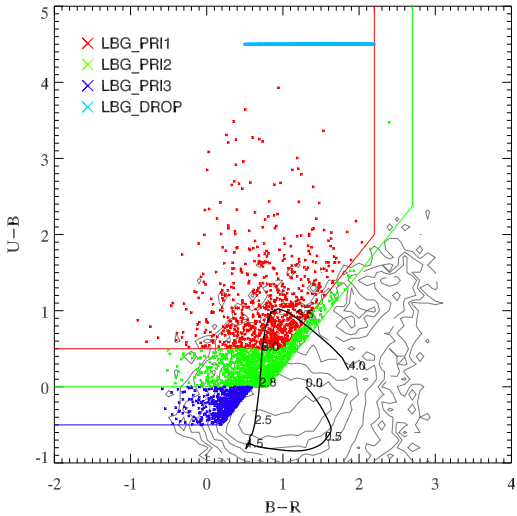}
\caption[$UBR$ colour-colour plot showing candidate selection in Q2359]{$UBR$ colour-colour plot showing candidate selection in Q2359. Objects selected as \lone, \ltwo, \lthree\ and \ldrop\ candidates are shown in different colours as indicated by the legend. The \ldrop\ candidates have been placed at $U-B=4.5$. The contours show the colour distribution of the rest of the objects in the field.}
\label{f-ubr-q23}
\end{figure}

\begin{figure}
\centering
\includegraphics[width=8.0cm]{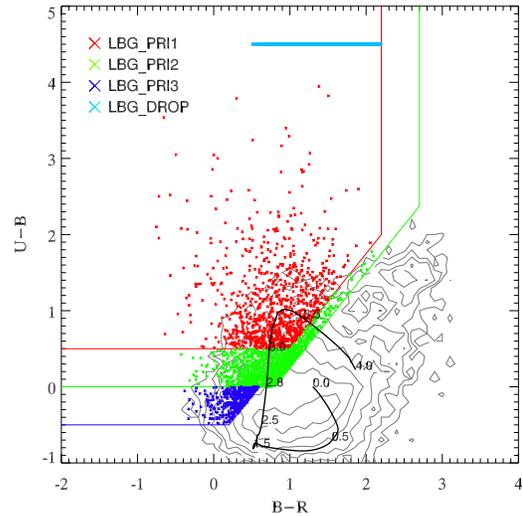}
\caption[$UBR$ colour-colour plot showing candidate selection in Q0301]{As for Fig.\ \ref{f-ubr-q23}, but for Q0301.}
\label{f-ubr-q03}
\end{figure}

\begin{figure}
\centering
\includegraphics[width=8.0cm]{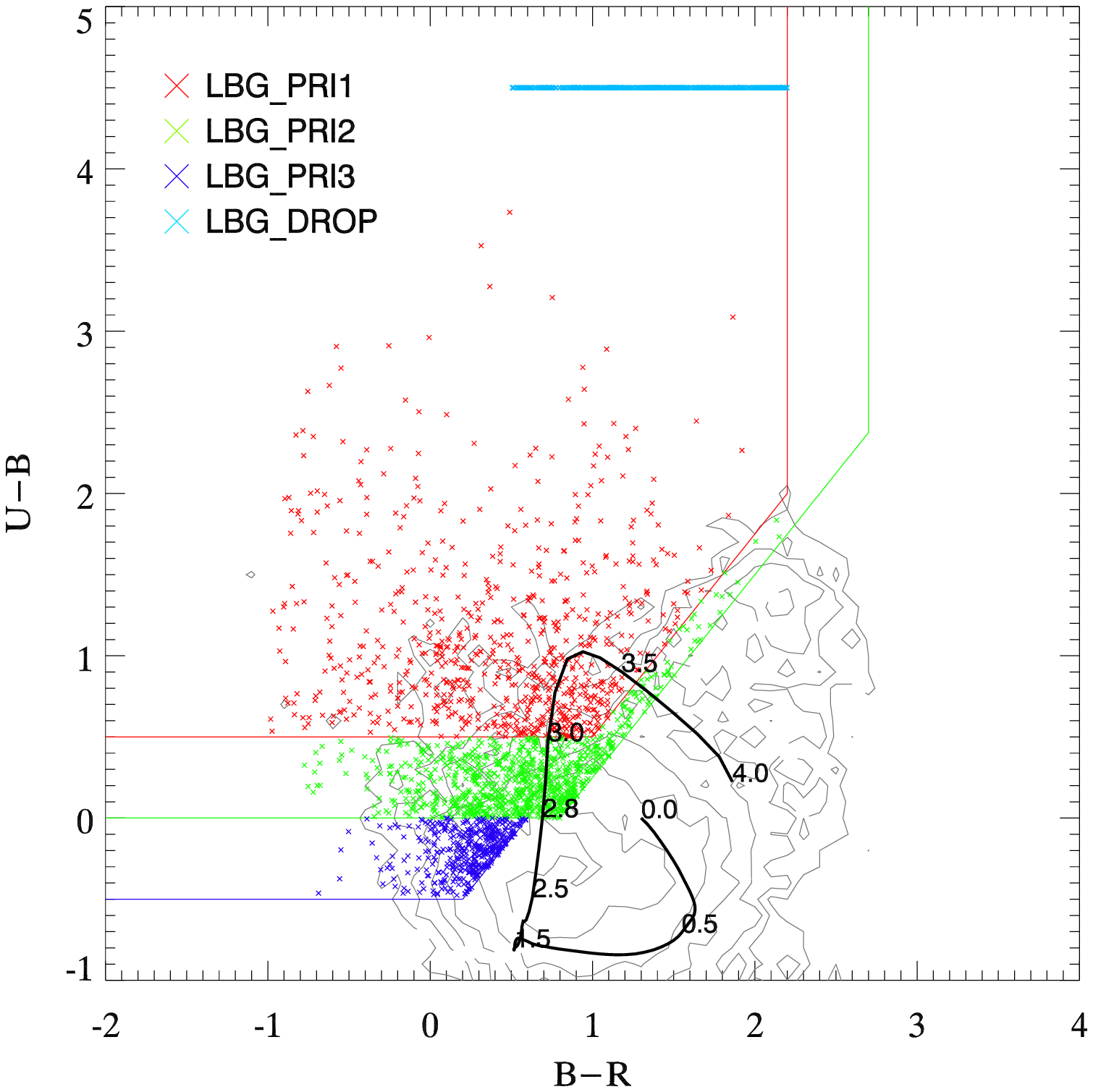}
\caption[$UBR$ colour-colour plot showing candidate selection in Q2231]{As for Fig.\ \ref{f-ubr-q23}, but for Q2231.}
\label{f-ubr-q22}
\end{figure}

Figs.\ \ref{f-ubr-q23}, \ref{f-ubr-q03} and \ref{f-ubr-q22} show $UBR$ colour--colour plots for Q2359, Q0301 and Q2231, respectively. In each plot, the \lone, \ltwo, \lthree\ and \ldrop\ candidates are indicated. A model colour--redshift track is also plotted, showing the expected evolutionary path of a star-forming galaxy from $z=0$ to $z=4$. This was produced using the Bruzual \& Charlot (2003) model, assuming a Chabrier IMF and an exponential SFR with e-folding time $\tau=9$ Gyr. The model indicates that our selection criteria (across all the priority groups) is predicted to isolate galaxies in the range $\approx2.5<z<3.8$. It also suggests that, of the sources that are confirmed as high-redshift LBGs, the \lthree{s} should typically be at a lower redshift than the \ltwo{s}, which in turn should be at lower redshift than the \lone{s}. \citetalias{2011MNRAS.414....2B} noted that this trend was detected in their LBG sample.

\subsubsection{$ugr$ selection}

\begin{figure}
\centering
\includegraphics[width=8.0cm]{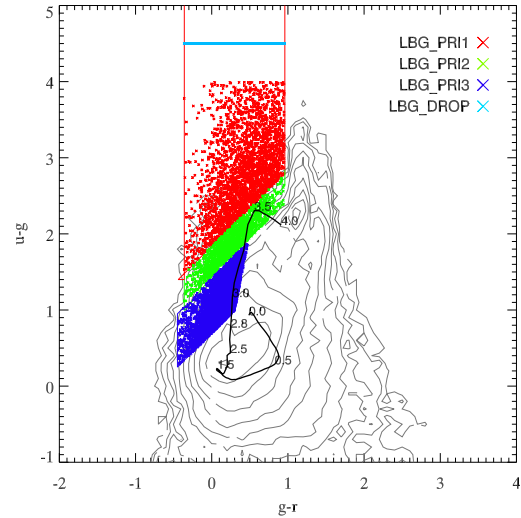}
\caption[$ugr$ colour-colour plot showing candidate selection in HE0940]{A $ugr$ colour-colour plot showing the selection of candidates in HE0940. Symbols are as in Fig.\ \ref{f-ubr-q23}.}
\label{f-ugr-he0}
\end{figure}

\begin{figure}
\centering
\includegraphics[width=8.0cm]{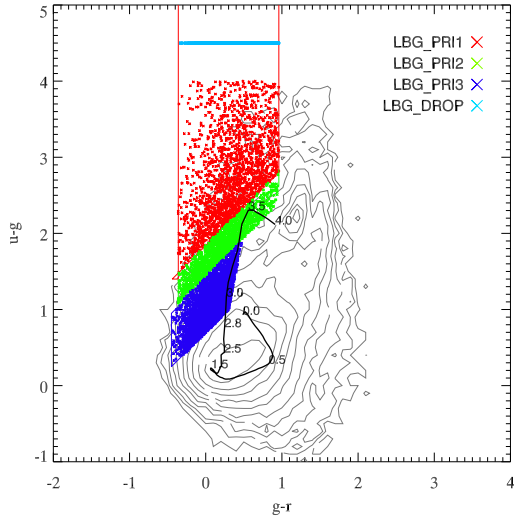}
\caption[$ugr$ colour-colour plot showing candidate selection in Q2348]{As for Fig.\ \ref{f-ugr-he0}, but for Q2348.}
\label{f-ugr-q48}
\end{figure}

In HE0940 and Q2348, LBG candidates were selected based on $ugr$ photometry. We have therefore adapted the criteria outlined above to account for the different colour bands. Again candidates were selected as either \lone, \ltwo, \lthree\ or \ldrop, defined as follows:

\begin{description}

\item[\lone] \hfill
\begin{itemize} 
\item $23<r\leq25$
\item $1.4<(u-g)<4.0$
\item $-0.36<(g-r)<0.96 $
\item $(g-r)<(u-g)-1.88 $
\end{itemize}

\item[\ltwo] \hfill
\begin{itemize}
\item $23<r\leq25 $
\item $(u-g)>1.0 $
\item $-0.36<(g-r)<0.96 $
\item $(g-r)<(u-g)-1.44 $
\item $\notin$ \lone
\end{itemize}

\item[\lthree] \hfill
\begin{itemize}
\item $23<r\leq25 $
\item $(g-r)<(u-g)-0.7 $
\item $-0.45<(g-r)<0.2(u-g)+0.1 $
\item $\notin$ \{\lone,\ltwo\}
\end{itemize}

\item[\ldrop] \hfill
\begin{itemize}
\item $23<r\leq25 $
\item $-0.36<(g-r)<0.96 $
\item No detection in $u$
\end{itemize}
\end{description}

Figs.\ \ref{f-ugr-he0} and \ref{f-ugr-q48} show the resulting candidates on $ugr$ colour-colour plots, and Table \ref{t-ncand} gives the numbers and sky densities of candidates in all 5 LBG fields.

\subsubsection{Comparing the LBG selections}

\begin{figure}
\centering
\includegraphics[width=8.0cm]{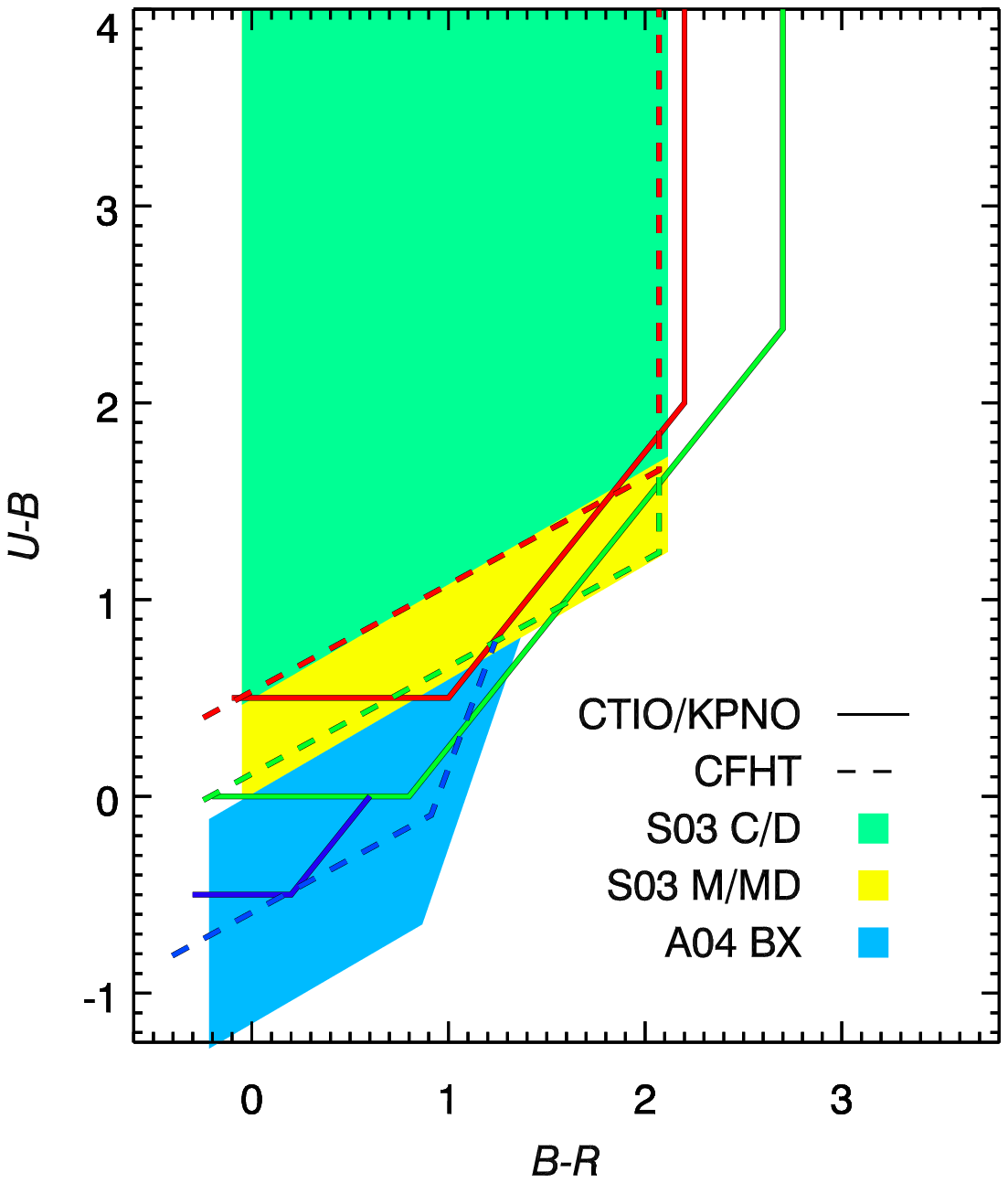}
\caption[]{Comparison of the $UBR$ and $ugr$ selections compared to the original \citet[][green and yellow filled regions]{steidel03} LBG and \citet[][cyan filled region]{2004ApJ...607..226A} BX $U_nGR$ selections. All three are presented in the Vega $UBR$ system.}
\label{fig:lbgselcomp}
\end{figure}

The $UBR$ and $ugr$ selections have been developed to mimic the LBG selection of \citet{steidel03} and the BX selection of \citet{2004ApJ...607..226A}. However, given the different sets of filters, the selection functions used here may not perfectly reproduce the original selections. For reference we show the selection functions used here overlayed on the original LBG and BX selections (transformed to the Vega $UBR$ system using the relations given by \citealt{1993AJ....105.2017S}) in Fig.~\ref{fig:lbgselcomp}. The CTIO/KPNO filter and CFHT MegaCAM filter (transformed to $UBR$ using the relations given on the CFHT website\footnote{http://www3.cadc-ccda.hia-iha.nrc-cnrc.gc.ca/megapipe/docs/filters.html} combined with those of \citealt{1996AJ....111.1748F}) selections are seen to agree well with the original \citet{steidel03} and \citet{2004ApJ...607..226A} selections. We note that we do not cover the entire of the \citet{2004ApJ...607..226A} BX region as doing so would bring a greater fraction of $z\lesssim2$ galaxies. In addition, our $UBR$ selection boundaries extend somewhat further in the positive $B-R$ extent. This region is populated by few galaxies in the $23<R\leq25$ range as evidenced by Figs.~\ref{f-ubr-q23}, \ref{f-ubr-q03} and \ref{f-ubr-q22}, but is included to catch faint galaxies that have been scattered in colour space due to photometric errors.

In terms of the resulting space densities, the LBG\_PRI1, LBG\_PRI2 and LBG\_DROP combined for the five fields give a mean space density of 2.00 arcmin$^{-2}$ for $R\leq25$, marginally higher than the combination of the C, D, M and MD LBG selections of \citealt{steidel03} that give a mean sky density of $\sim1.8$ arcmin$^{-2}$ for ${\cal R}<25.5$. Taking the LBG\_PRI3 candidates, we obtain a mean sky density of 0.37 arcmin$^{-2}$ in the $UBR$ fields and 1.38 arcmin$^{-2}$ in the $ugr$ fields. The LBG\_PRI3 selection is intended to provide additional galaxies at $2<z<2.5$ and overlaps to some extent with the \citet{2004ApJ...607..226A} BX selection (as illustrated in Fig.~\ref{fig:lbgselcomp}). As expected the number densities here for the LBG\_PRI3 selection are much lower than the BX selection, which obtains numbers of $\sim5$ arcmin$^{-2}$ at ${\cal R}<25.5$, due to only sampling a subset of the BX selection.


\begin{table*}
\centering
\caption[Numbers and sky densities of LBG candidates]{Numbers and sky densities of \zth\ LBG candidates in each of the fields. For each priority class, the first column shows the total number of candidates selected and the second, italicised column shows the density in arcmin$^{-2}$. The last two columns show the figures for all \zth\ LBG candidates.}
\label{t-ncand}
\begin{tabular}{lcccccccccc}
\\
\hline
Field & \multicolumn{2}{c}{\lone} & \multicolumn{2}{c}{\ltwo} & \multicolumn{2}{c}{\lthree} & \multicolumn{2}{c}{\ldrop} & \multicolumn{2}{c}{All} \\
\hline
Q2359+0653  & 795   & \emph{0.61} & 1,130 & \emph{0.87} & 549   & \emph{0.42} &   709 & \emph{0.55} & 3,183  & \emph{2.45} \\
Q0301-0035  & 891   & \emph{0.69} & 1,227 & \emph{0.95} & 433   & \emph{0.33} & 1,014 & \emph{0.78} & 3,565  & \emph{2.75} \\
Q2231+0015  & 748   & \emph{0.65} & 948   & \emph{0.82} & 424   & \emph{0.37} &   514 & \emph{0.45} & 2,634  & \emph{2.29} \\
HE0940-1050 & 2,657 & \emph{0.69} & 1,896 & \emph{0.49} & 5,370 & \emph{1.40} & 3,663 & \emph{0.96} & 13,586 & \emph{3.54} \\
Q2348-011   & 1,843 & \emph{0.52} & 1,624 & \emph{0.46} & 4,808 & \emph{1.35} & 1,850 & \emph{0.52} & 10,125 & \emph{2.84} \\
\hline 
\\
\end{tabular}
\end{table*}

\subsection{Spectroscopy}

\subsubsection{Observations} \label{ss-specobs}

The LBG candidates were targeted in spectroscopic follow-up observations with the VLT VIMOS spectrograph between September 2008 and December 2009, with programme IDs 081.A-0418(B) (Q2231), 081.A-0418(D) (Q2359), 081.A-0418(F) (Q0301), 082.A-0494(B) (HE0940) and 082.A-0494(D) (Q2348). The observations were done during dark time in generally good conditions with a typical seeing of $\approx1$\arcsec\ and an airmass of 1.0--1.3. Table \ref{t-specobs} gives details of all the fields observed.

The VIMOS instrument \citep{2003SPIE.4841.1670L} comprises four separate CCDs, each with a field of view of $7\marcmin \times 8\marcmin$. These four arms are arranged in a $2\times2$ grid with a $\approx2\marcsec$ gap between each CCD, giving a total $16\marcmin \times 18\marcmin$ FoV as quoted previously. Of this 288 arcmin$^2$ field, 224 arcmin$^2$ is covered by the detector.

Our observations utilised the low resolution blue (LR\_Blue) grism and the order sorting blue (OS\_Blue) filter, resulting in a wavelength range of 3700\AA--6700\AA, blazing at $\approx$4000\AA. This wavelength range is ideal for our survey, detecting the \lya\ line at $2.0<z<4.5$. The resolving power of the spectrograph in this configuration is $R=180$ assuming a 1\arcsec\ slit (as used in these observations), which gives a resolution element of $\Delta\lambda\approx22$\AA\ at the blaze wavelength. The spectral dispersion is 5.3\AA/pixel.

\begin{table*}
\centering
\caption{Details of spectroscopic observations for the LBG target fields presented in this paper.}
\label{t-specobs}
\begin{tabular}{cccccccc}
\\
\hline
Field & Subfield & RA$^a$ & Dec$^a$ & Exposure$^b$ & Airmass & Seeing & Dates\\
\hline
Q2359   & f1 & 00:01:09.94 & +07:03:26.8 & 10 & $1.1-1.2$ & $0.6-1.7$\arcsec\ & 23--25 Sep 2008\\
Q2359   & f2 & 00:01:12.92 & +07:16:39.2 & 10 & $1.1-1.3$ & $0.6-1.3$\arcsec\ & 3, 20--21 Oct 2008\\
Q2359   & f3 & 00:02:11.50 & +07:15:33.9 & 10 & $1.2-1.4$ & $0.6-1.3$\arcsec\ & 3, 21--25 Nov 2008\\
Q2359   & f4 & 00:02:12.89 & +07:02:19.1 & 10 & $1.2-1.3$ & $0.5-1.1$\arcsec\ & 26--30 Nov 2008\\
\hline
Q2231   & f1 & 22:34:28.19 & --00:06:03.3 & 10 & $1.1-1.2$ & $0.5-1.0$\arcsec\ & 23, 28 Oct 2008\\
Q2231   & f2 & 22:34:28.55 & +00:06:13.2 & 10 & $1.1-1.2$ & $0.4-0.9$\arcsec\ & 21--22 Oct 2008\\
Q2231   & f3 & 22:33:39.51 & --00:06:10.8 & 10 & $1.1-1.2$ & $0.5-1.0$\arcsec\ & 3 Aug; 27 Jul 2008\\
\hline
Q0301   & f1 & 03:04:20.12 & --00:14:28.8 & 12 & $1.1-1.3$ & $0.7-1.2$\arcsec\ & 23, 31 Oct 2008\\
Q0301   & f2 & 03:03:10.27 & --00:16:18.7 & 12 & $1.1-1.2$ & $0.7-1.5$\arcsec\ & 21--23 Nov 2008\\
Q0301   & f3 & 03:03:15.41 & --00:30:40.0 & 12 & $1.1-1.2$ & $0.7-1.5$\arcsec\ & 25--26 Nov 2008\\
Q0301   & f4 & 03:04:15.56 & --00:28:59.1 & 12 & $1.1-1.2$ & $0.7-1.2$\arcsec\ & 24 Sep; 1, 7 Oct 2008\\
\hline
HE0940 & f4 & 09:42:10.00 & --10:54:30.3 & 11.2 & $1.0-1.3$ & $0.8-1.5$\arcsec\ & 1 Feb 2009\\
HE0940 & f5 & 09:43:07.47 & --11:24:50.3 & 11.2 & $1.0-1.3$ & $0.7-1.4$\arcsec\ & 3 Feb 2009\\
HE0940 & f6 & 09:41:59.99 & --11:24:50.4 & 11.2 & $1.0-1.2$ & $0.5-1.2$\arcsec\ & 20--21 Feb 2009\\
HE0940 & f7 & 09:44:14.99 & --11:24:49.9 & 11.2 & $1.0-1.2$ & $0.5-1.3$\arcsec\ & 22, 24 Feb 2009\\
HE0940 & f8 & 09:43:21.49 & --10:41:00.5 & 11.2 & $1.0-1.2$ & $0.5-1.0$\arcsec\ & 26--27 Feb 2009\\
HE0940 & f9 & 09:42:09.99 & --10:40:59.8 & 11.2 & $1.0-1.3$ & $0.5-1.2$\arcsec\ & 2 Feb 2009\\
\hline
Q2348   & f1 & 23:51:50.08 & --00:54:21.9  & 11.5 & $1.1-1.2$ & $0.5-1.7$\arcsec\ & 23--25 Jul 2009\\
Q2348   & f2 & 23:50:45.09 & --00:54:22.2 & 11.5 & $1.0-1.1$ & $0.5-1.0$\arcsec\ & 19--20 Jul 2009\\
Q2348   & f3 & 23:49:40.07 & --00:54:22.6 & 11.5 & $1.0-1.2$ & $0.4-0.8$\arcsec\ & 27 Jul 2009\\
Q2348   & f4 & 23:51:50.12 &  --00:37:31.6 & 11.5 & $1.1-1.2$ & $0.5-1.5$\arcsec\ & 20--21 Aug 2009\\
Q2348   & f5 & 23:50:45.05 &  --00:37:31.5 & 11.5 & $1.0-1.2$ & $0.7-1.4$\arcsec\ & 16--20 Sep 2009\\
Q2348   & f6 & 23:49:40.00 &  --00:37:32.0 & 11.5 & $1.1-1.2$ & $0.8-1.3$\arcsec\ & 24--25 Sep 2009\\
Q2348   & f7 & 23:51:50.12 &  --01:07:31.4 & 11.5 & $1.1-1.2$ & $0.7-1.0$\arcsec\ & 12, 20 Oct 2009\\
Q2348   & f8 & 23:50:45.00 &  --01:07:32.0 & 11.5 & $1.1-1.3$ & $0.7-1.3$\arcsec\ & 22 Nov, 10 Dec 2009\\
Q2348   & f9 & 23:49:40.00 &  --01:07:32.0 & 11.5 & $1.1-1.3$ & $0.8-1.5$\arcsec\ & 15--22 Nov 2009\\
\hline
\multicolumn{7}{l}{$^a$ J2000 coordinates of subfield centre}\\
\multicolumn{7}{l}{$^b$ in ks}\\
\\
\end{tabular}
\end{table*} 

The slit masks were designed using the \textsc{vmmps} software which is standard for VIMOS observations. The aims for mask design are (a) to maximise the number of observed targets, (b) to favour higher-priority targets and (c) to ensure slits are of sufficient size to allow a robust sky subtraction. Since these aims are frequently in conflict with one another, the mask design process is one of attempting to optimise the observations to satisfy all three as well as possible. Point (c) is addressed by setting a minimum slit length of 8\arcsec\ (40 pixels given the pixel scale of 0.205\arcsec/pixel). Slit length was increased as much as possible where such an increase would not prevent the observation of an additional target --- that is, where it did not conflict with point (a). Finally, in order to optimise slit allocation, some targets were added to fill gaps that fulfilled the given selection criteria, but with fainter $R$ magnitudes down to a limit of $R=25.5$.

With the LR\_Blue grism, each spectrum spans 640 pixels along the dispersion axis. Assuming a 40 pixel slit width as specified above, this would allow for a possible total of over 300 slits on the full 4k$\times$2k detector. This is however not practically achievable given the density of LBG candidates, and is hampered further by the need to select high-priority candidates (point b), which have an even lower sky density. Our final slit masks therefore typically contain some 50--70 slits per quadrant.

\subsubsection{Data reduction}

The spectroscopic data have all been reduced using the VIMOS \textsc{esorex} reduction pipeline. Using bias frames, flat fields and arc lamp exposures taken for each mask during each observing run, the pipeline generates bias-subtracted, flat-fielded, wavelength-calibrated science frames consisting of a series of 2D spectra. Following \citetalias{2011MNRAS.414....2B} we use the \texttt{imcombine} procedure in \textsc{iraf} to combine the reduced frames from each observing block, generating a master science frame for each quadrant of each field. When combining the frames we use the \texttt{crreject} mode, designed to remove cosmic rays by rejecting pixels with significant positive spikes. We have also used the \texttt{avsigclip} rejection mode with a rejection threshold of $3\sigma$, and find that our results are not significantly affected, suggesting that our results do not depend strongly on the parameters used to combine the science frames at this stage.

We extract 1D spectra from the reduced, combined 2D spectra using the \textsc{idl} routine \texttt{specplot}. One-dimensional object and sky spectra are found by averaging across the respective apertures, and the sky spectrum is then subtracted from the object spectrum to give a final spectrum for the object.

In some cases there remain significant sources of contamination in the final object spectrum. These can arise from bad pixels, either in the object or sky aperture, from contamination from the zeroth order from other slits, or more frequently from the bright sky emission lines [O\,\textsc{i}] 5577\AA, [Na\,\textsc{i}] 5990\AA\ and [O\,\textsc{i}] 6300\AA; in either case, the resulting contamination may manifest itself as either a positive or a negative spike in the spectrum. Such artefacts are, however, easily spotted during a routine inspection of the 2D spectrum.

\subsection{Identification of targets} \label{s-targetid}

Every source targeted for spectroscopic observation is inspected visually, in both the 2D and 1D spectra, to determine where possible a redshift and classification. Sources are classified as either \zth\ Lyman-break galaxies, low-redshift galaxies, QSOs or Galactic stars. The LBGs are divided into those showing \lya\ emission (designated LBe) and those showing \lya\ absorption (LBa). QSOs are determined by the presence of typical AGN emission features, in particular Ly$\alpha$ and C\,\textsc{IV}. Stars are classified by comparison to template spectra: in particular we check for A, F, G, K and M stars. 

In determining the redshift and classification the spectral feature primarily used in the case of LBGs is the \lya\ emission/absorption line at 1216\AA; for lower redshift galaxies it is the [O\,\textsc{ii}] emission line at 3727\AA. In addition to these, some of the following features are used: 

\noindent For \zth\ LBGs:
\begin{itemize}
\item Lyman limit, 912\AA;
\item Ly$\beta$ emission/absorption, 1026\AA\;
\item O\,\textsc{VI} 1032\AA, 1038\AA;
\item Ly$\alpha$ forest, $<$1215.67\AA;
\item Ly$\alpha$ emission/absorption, 1215.67\AA;
\item Inter-stellar medium (ISM) absorption lines:
\begin{itemize}
\item Si\,\textsc{II} 1260.4\AA;
\item O\,\textsc{I}$+$Si\,\textsc{II} 1303\AA;
\item C\,\textsc{II} 1334\AA;
\item Si\,\textsc{IV} doublet 1393\AA\ \& 1403\AA;
\item Si\,\textsc{II} 1527\AA;
\item C\,\textsc{IV} doublet absorption, 1548-1550\AA.
\item Fe\,\textsc{II} 1608\AA;
\item Al\,\textsc{II} 1670\AA;
\end{itemize}
\end{itemize}

\noindent For low-$z$ galaxies:
\begin{itemize}
\item CN absorption 3833\AA;
\item K-band absorption 3934\AA;
\item HK break 4000\AA;
\item H$\delta$ emission 4102\AA;
\item H$\beta$ emission/absorption 4861\AA;
\item O\,\textsc{iii} emission 4959\AA;
\item O\,\textsc{iii} emission 5007\AA;
\end{itemize}

The presence of the HK break causes these interlopers to appear fairly frequently in our spectroscopic samples, since these features mimic the Lyman break on which our selection is based. The ISM absorption features listed above for LBGs are therefore of considerable importance in identifying genuine \zth\ galaxies. For every target which is identified, we assign a quality parameter to the redshift determined, in the range $0\le Q \le1$. A quality of $Q\le 0.4$ indicates that a possible redshift has been determined, but is not considered a robust measurement. Above this, for LBGs, the quality parameters indicate that the redshift is based on the following features:

\begin{itemize}
\item $Q=0.5$ --- a spectral break with some weak \lya\ emission/absorption and low-SNR ISM absorption features, or strong \lya\ emission but with no detected continuum \
\item $Q=0.6$ --- a spectral break with high-SNR \lya\ emission/absorption plus low-SNR ISM absorption features \
\item $Q=0.7$ --- a spectral break with high-SNR \lya\ emission/absorption plus unambiguous, high-SNR ISM absorption features \
\item $Q=0.8$ --- a spectral break with high-SNR \lya\ emission/absorption plus high-SNR absorption and lower-SNR emission lines (e.g. Si\,\textsc{ii} 1265\AA, 1309\AA; He\,\textsc{ii} 1640\AA) \
\item $Q=0.9$ --- as for Q=0.8, but reserved for highest signal-to-noise objects only \
\end{itemize}

\begin{table*}
\centering\caption[]{Example LBG identifications in the Q2359+0653.}\label{tab:lbgQ2359+0653}
\begin{tabular}{lcccccccc}
\hline
                          ID &            R.A. &            Dec. &    $U-B$ &    $B-R$ &      $R$ &  $z_{Ly\alpha}$ &       $z_{ISM}$ & $Q_{ID}$\\
                             &       \multicolumn{2}{c}{(J2000)} &          &          &          &                 &                 &         \\
\hline
   VLRS J000139.85+070221.66 &       0.4160563 &       7.0393505 &     0.63 &     0.76 &  23.5500 &          2.4762 &          2.4682 &      0.5\\
   VLRS J000133.54+070127.57 &       0.3897395 &       7.0243263 &     0.13 &     0.15 &  25.3600 &          2.5707 &          2.5603 &      0.5\\
   VLRS J000118.84+070106.55 &       0.3285175 &       7.0184855 &     1.29 &     1.21 &  23.8900 &          3.0374 &          3.0294 &      0.5\\
   VLRS J000131.05+070106.56 &       0.3793770 &       7.0184898 &     0.58 &     0.65 &  25.3900 &          2.7910 &          2.7967 &      0.5\\
   VLRS J000141.27+070106.35 &       0.4219468 &       7.0184293 &     1.51 &     0.12 &  24.3800 &          2.6508 &          2.6428 &      1.0\\
   VLRS J000140.69+070044.11 &       0.4195270 &       7.0122533 &     0.47 &     0.65 &  23.7800 &          2.8162 &          2.8082 &      0.5\\
                         ... &             ... &             ... &      ... &      ... &      ... &             ... &             ... &      ...\\
                         ... &             ... &             ... &      ... &      ... &      ... &             ... &             ... &      ...\\
\hline
\end{tabular}
\end{table*}
\begin{table*}
\centering\caption[]{Example LBG identifications in the Q0301-0035.}\label{tab:lbgQ0301-0035}
\begin{tabular}{lcccccccc}
\hline
                          ID &            R.A. &            Dec. &    $U-B$ &    $B-R$ &      $R$ &  $z_{Ly\alpha}$ &       $z_{ISM}$ & $Q_{ID}$\\
                             &       \multicolumn{2}{c}{(J2000)} &          &          &          &                 &                 &         \\
\hline
   VLRS J030434.85-001549.27 &      46.1452103 &      -0.2636854 &     0.59 &     0.61 &  24.4700 &          2.6132 &          2.6041 &      0.7\\
   VLRS J030435.40-001607.15 &      46.1474953 &      -0.2686527 &     1.56 &     1.45 &  23.4100 &          2.5969 &          2.6157 &      0.7\\
   VLRS J030439.49-001619.35 &      46.1645317 &      -0.2720422 &     1.08 &     0.62 &  24.5100 &          2.9570 &          2.9490 &      0.5\\
   VLRS J030438.22-001647.63 &      46.1592560 &      -0.2798966 &    -0.22 &     0.37 &  23.8600 &          2.7098 &          2.7292 &      0.6\\
   VLRS J030435.86-001654.14 &      46.1494179 &      -0.2817046 &     0.80 &     0.92 &  25.0700 &          2.8887 &          2.8807 &      1.0\\
   VLRS J030426.38-001701.38 &      46.1099281 &      -0.2837157 &     0.21 &     0.56 &  24.4900 &          2.4651 &          2.4571 &      0.6\\
                         ... &             ... &             ... &      ... &      ... &      ... &             ... &             ... &      ...\\
                         ... &             ... &             ... &      ... &      ... &      ... &             ... &             ... &      ...\\
\hline
\end{tabular}
\end{table*}
\begin{table*}
\centering\caption[]{Example LBG identifications in the HE0940-1050.}\label{tab:lbgHE0940-1050}
\begin{tabular}{lcccccccc}
\hline
                          ID &            R.A. &            Dec. &    $u-g$ &    $g-r$ &      $r$ &  $z_{Ly\alpha}$ &       $z_{ISM}$ & $Q_{ID}$\\
                             &       \multicolumn{2}{c}{(J2000)} &          &          &          &                 &                 &         \\
\hline
   VLRS J094225.83-105744.50 &     145.6076355 &     -10.9623623 &     2.57 &     0.35 &  23.6400 &          2.8804 &          2.8810 &      0.5\\
   VLRS J094240.69-105753.44 &     145.6695251 &     -10.9648447 &     1.88 &     0.35 &  24.1100 &          3.1456 &          3.1376 &      0.5\\
   VLRS J094220.01-105900.05 &     145.5833588 &     -10.9833469 &      --- &    -0.14 &  24.3800 &          2.2010 &          2.1930 &      0.5\\
   VLRS J094217.39-105923.95 &     145.5724640 &     -10.9899855 &      --- &     0.16 &  23.8500 &          2.5153 &          2.5073 &      0.5\\
   VLRS J094217.51-105935.92 &     145.5729675 &     -10.9933100 &      --- &     0.79 &  24.2600 &          2.8139 &          2.8119 &      0.6\\
   VLRS J094242.29-110121.16 &     145.6762085 &     -11.0225439 &     0.74 &    -0.03 &  23.9900 &          2.4613 &          2.4595 &      0.5\\
                         ... &             ... &             ... &      ... &      ... &      ... &             ... &             ... &      ...\\
                         ... &             ... &             ... &      ... &      ... &      ... &             ... &             ... &      ...\\
\hline
\end{tabular}
\end{table*}
\begin{table*}
\centering\caption[]{Example LBG identifications in the Q2231+0015.}\label{tab:lbgQ2231+0015}
\begin{tabular}{lcccccccc}
\hline
                          ID &            R.A. &            Dec. &    $U-B$ &    $B-R$ &      $R$ &  $z_{Ly\alpha}$ &       $z_{ISM}$ & $Q_{ID}$\\
                             &       \multicolumn{2}{c}{(J2000)} &          &          &          &                 &                 &         \\
\hline
   VLRS J223439.00+000341.29 &     338.6625061 &       0.0614693 &      --- &     0.72 &  24.1500 &          2.8428 &          2.8291 &      0.6\\
   VLRS J223459.87+000308.07 &     338.7494507 &       0.0522424 &     1.01 &     0.78 &  23.7400 &          2.4879 &          2.4789 &      0.5\\
   VLRS J223450.20+000232.38 &     338.7091675 &       0.0423284 &      --- &     1.92 &  23.7800 &          2.8934 &          2.8927 &      0.7\\
   VLRS J223459.03+000051.79 &     338.7459717 &       0.0143855 &      --- &     0.54 &  24.3200 &          2.8037 &          2.7957 &      0.9\\
   VLRS J223442.76-000028.59 &     338.6781616 &      -0.0079414 &     0.32 &     1.01 &  23.6900 &          2.1897 &          2.1817 &      0.5\\
   VLRS J223447.81-000041.63 &     338.6991882 &      -0.0115636 &      --- &     0.87 &  24.8000 &          2.8874 &          2.8735 &      0.5\\
                         ... &             ... &             ... &      ... &      ... &      ... &             ... &             ... &      ...\\
                         ... &             ... &             ... &      ... &      ... &      ... &             ... &             ... &      ...\\
\hline
\end{tabular}
\end{table*}
\begin{table*}
\centering\caption[]{Example LBG identifications in the Q2348-011.}\label{tab:lbgQ2348-011}
\begin{tabular}{lcccccccc}
\hline
                          ID &            R.A. &            Dec. &    $u-g$ &    $g-r$ &      $r$ &  $z_{Ly\alpha}$ &       $z_{ISM}$ & $Q_{ID}$\\
                             &       \multicolumn{2}{c}{(J2000)} &          &          &          &                 &                 &         \\
\hline
   VLRS J235206.92-005646.70 &     358.0288391 &      -0.9463067 &     1.30 &     0.30 &  24.4600 &          3.1430 &          2.3707 &      0.5\\
   VLRS J235200.98-005903.11 &     358.0040894 &      -0.9841969 &     2.50 &     0.46 &  24.2400 &          3.3532 &          3.3435 &      0.5\\
   VLRS J235201.68-010002.18 &     358.0069885 &      -1.0006067 &     1.23 &     0.31 &  24.9700 &          2.8632 &          2.8603 &      0.5\\
   VLRS J235155.14-010104.04 &     357.9797363 &      -1.0177902 &     1.66 &     0.15 &  23.8600 &          2.7523 &          2.7503 &      0.7\\
   VLRS J235209.28-005535.49 &     358.0386658 &      -0.9265237 &     1.28 &     0.13 &  24.0700 &          2.6274 &          2.6194 &      0.7\\
   VLRS J235202.62-004747.89 &     358.0109253 &      -0.7966368 &     2.11 &     0.32 &  24.2700 &          3.0640 &          3.0646 &      0.5\\
                         ... &             ... &             ... &      ... &      ... &      ... &             ... &             ... &      ...\\
                         ... &             ... &             ... &      ... &      ... &      ... &             ... &             ... &      ...\\
\hline
\end{tabular}
\end{table*}

\subsection{LBG sample}

\subsubsection{Sky densities, redshift distributions and completeness}

\begin{figure*}
\centering
\includegraphics[width=0.8\textwidth]{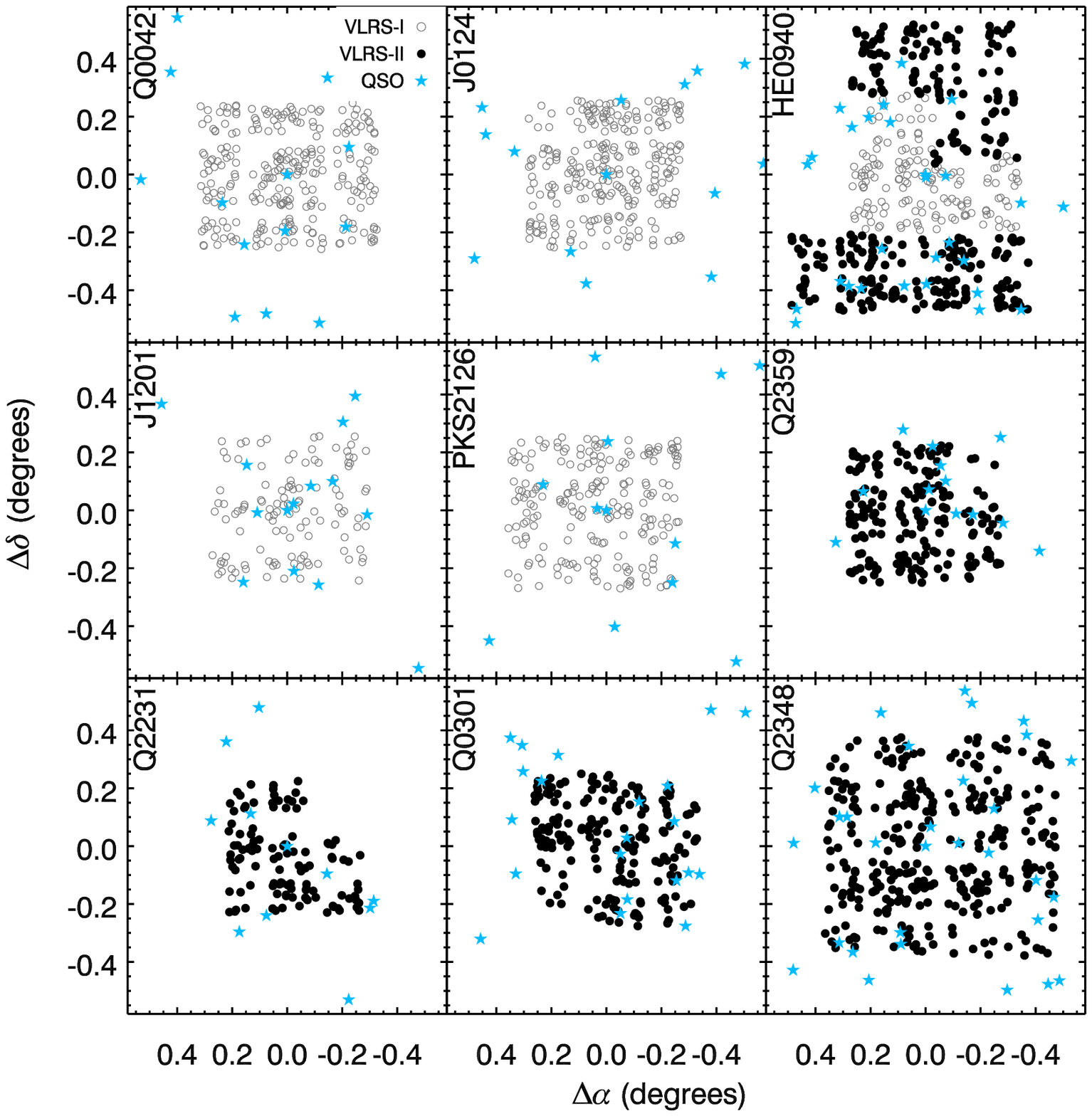}
\caption[Sky distribution]{Sky distribution of LBGs and QSOs in the VLRS fields. Grey open circles show LBGs presented in \citetalias{2011MNRAS.414....2B}, black filled circles show LBGs identified in the current work and cyan stars show known QSOs.}
\label{fig:fields}
\end{figure*}

In total, the VLRS now consists of \totlbg\ spectroscopically confirmed LBGs in $\approx$ 10,000 arcmin$^2$ (a density of $\approx$ 0.20 arcmin\psq). \numlbg\ of these are within a magnitude limit of $R\leq25$, whilst the remainder form a non-uniform sample of $25<R\leq25.5$ LBGs that were observed as part of optimising slit allocations in the spectroscopic observations. We show the sky distribution of LBGs from both \citetalias[][ (open grey circles)]{2011MNRAS.414....2B} and this paper (filled black circles) for all nine VLRS fields in Fig.~\ref{fig:fields}. Known QSOs in the fields are also plotted (cyan stars). The total numbers of $R\leq25$ $Q\ge0.5$ sources identified in each of the 5 fields presented here are given in Table \ref{t-ids}. We present examples of the first six LBGs in each of the fields in Tables~\ref{tab:lbgQ2359+0653} to \ref{tab:lbgQ2348-011}. The full tables will be made available online at http://star-www.dur.ac.uk/$\sim$bielby/vlrs/.

For the $z>2$ sample, the VLRS now consists of 944 $Q=0.5$, 492 $Q=0.6$, 318 $Q=0.7$, 147 $Q=0.8$ and 93 $Q=0.9$ galaxies at a magnitude limit of $R\leq25$.

\begin{table*}
\centering
\caption{Numbers of objects in each target field spectroscopically identified at $Q\ge0.5$ and $R\leq25$ as either high-$z$ LBGs, $z<2$ galaxies, AGN/QSOs, stars or unidentified.}
\label{t-ids}
\begin{tabular}{lcccc}
\\
\hline
Field & \zth\ LBGs & $z<2$ galaxies & QSO/AGN & Stars \\
\hline
Q2359    & 143 (0.18 arcmin\psq)   & 67  & 5  & 8  \\
Q0301    & 164 (0.21 arcmin\psq)   & 61  & 10 & 13 \\
Q2231    & 108 (0.18 arcmin\psq)   & 80  & 6  & 18 \\
HE0940   & 358 (0.30 arcmin\psq)   & 186 & 4  & 48 \\
Q2348    & 303 (0.17 arcmin\psq)   & 100 & 11 & 34 \\
\hline
Total    & 1,076 (0.21 arcmin\psq) & 494 & 36 & 121   \\
\hline
\\
\end{tabular}
\end{table*}

\begin{figure}
\centering
\includegraphics[width=8.0cm]{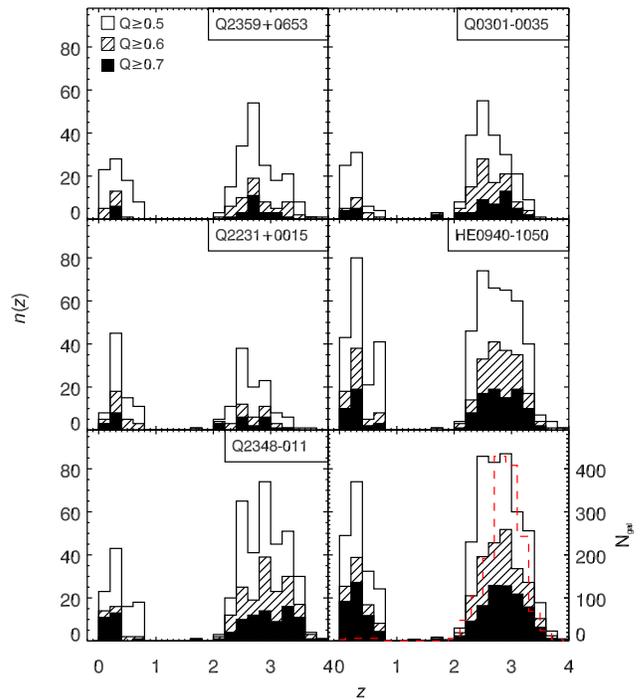}
\caption{Redshift distributions of identified sources in each of the target fields (first five panels) and for the whole sample of \numlbg\ $R\leq25$ galaxies (bottom right panel). In each panel we show the distribution for the full sample ($Q\ge0.5$), as well as for the $Q\ge0.6$ (hashed) and $Q\ge0.7$ (filled) subsets, where $Q$ is the redshift quality parameter. In the final panel, we also show the redshift distribution of the \citet{steidel03} galaxies used in this work (dashed red line).}
\label{f-nzqid5}
\end{figure}

Fig.\ \ref{f-nzqid5} shows the $n(z)$ distributions of all sources with measured redshifts in each of the 5 LBG fields. The figure shows that LBGs in HE0940 and Q2348, where LBGs were selected in $ugr$, have higher average redshifts than in the $UBR$-selected fields, suggesting that the $ugr$ criteria bias the selection toward higher $z$. It is also notable from Table \ref{t-ids} that the $ugr$ selection appears to include more Galactic stars. Future $ugr$-selected LBG surveys may wish to alter our colour criteria to better avoid stellar interlopers.

Fig.\ \ref{f-nzqid5} also shows $n(z)$ for the subsets of sources with $Q\ge0.6$ and $Q\ge0.7$. We note that in any given field, the \emph{distributions} of sources at $Q\ge0.5$, $Q\ge0.6$ or $Q\ge0.7$ are approximately the same --- the LBGs with higher ID qualities are not skewed to lower or higher redshift, for example --- suggesting that the redshift distributions shown are fairly robust. The average redshifts and standard deviations are given in Tab.~\ref{t-lbgstats}. The redshift distribution of the full LBG sample has a mean redshift of $\meanz\pm0.01$ and a standard deviation of 0.34, and is shown in the lower right panel of Fig.\ \ref{f-nzqid5}.


\begin{table}
\centering
\caption[Redshift distribution statistics for spectroscopically confirmed LBGs]{Redshift distribution statistics for spectroscopically confirmed, $R\leq25$ $Q\ge0.5$ LBGs in the 5 observed fields. In each case the mean redshift $\bar{z}$ (with standard error), median redshift $\tilde{z}$ and standard deviation $\sigma$ of the distribution is given.}
\label{t-lbgstats}
\begin{tabular}{ccccccccccc}
\\
\hline
Field & $\bar{z}$ & $\tilde{z}$ & $\sigma$ \\
\hline
Q2359   & $2.81\pm0.03$ & $2.74$ & $0.36$ \\ 
Q0301   & $2.64\pm0.02$ & $2.59$ & $0.31$ \\ 
Q2231   & $2.68\pm0.03$ & $2.65$ & $0.30$ \\
HE0940  & $2.79\pm0.02$ & $2.77$ & $0.34$ \\
Q2348   & $2.90\pm0.02$ & $2.92$ & $0.36$ \\
\hline 
Total   & $\meanz\pm0.01$ & $2.76$ & $0.35$ \\
\hline
\\
\end{tabular}
\end{table}

\begin{figure}
\centering
\includegraphics[width=8.0cm]{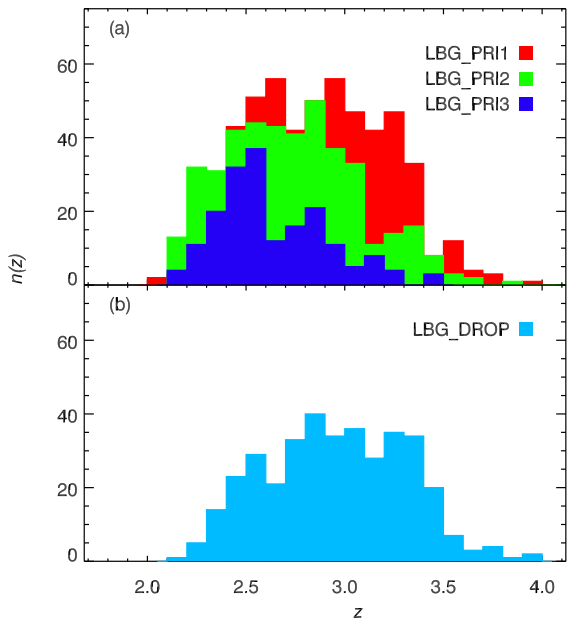}
\caption[Redshift distributions of identified sources separated by candidate priority class]{Redshift distribution of all identified sources our 5 target fields, separated by initial candidate priority. Panel \emph{(a)} shows the \lone, \ltwo\ and \lthree\ classes, while the \ldrop{s} are shown separately in panel \emph{(b)} for clarity. Colours are as in Figs.\ \ref{f-ubr-q23}--\ref{f-ugr-he0}.}
\label{f-nzpri}
\end{figure}

Fig.~\ref{f-nzpri} shows, as anticipated in \S\ref{s-candsel}, that candidates selected as \lone\ lie at higher redshift than the \ltwo\ candidates, which are in turn at higher redshift than \lthree{s}. Quantitatively, we find that the \lone{s} have a mean redshift of $\bar{z}=2.83\pm0.02$, the \ltwo{s} have $\bar{z}=2.71\pm0.02$ and the \lthree{s} have $\bar{z}=2.61\pm0.02$. The \ldrop\ candidates are shown in a separate panel for clarity, and have the highest mean redshift of all the groups, with $\bar{z}=2.93\pm0.02$.

In total, for the $R\leq25$ targets, we make successful identifications in $\approx30\%$ of VIMOS slits and of those identified, $62\%$ are identified as $z>2$ galaxies. In terms of the unidentified fraction, these are likely predominantly faint $z>2$ galaxies (most likely dominated by dusty, absorbed galaxies with no Ly$\alpha$ emission) and relatively featureless $1\lesssim z\lesssim 2$ galaxies. We note that the $38\%$ contamination rate is somewhat higher than that quoted for the \citet{steidel03} and subsequent samples. This is in part likely the result of the shallower depths and differing filters used in the colour selections. Additionally, following the results of other authors \citep[e.g.][]{2008ApJS..175...48R}, it is likely that the faint population that has avoided identification in our observations is less prone to contamination and as such likely has a higher percentage of $z\sim3$ galaxies than the $62\%$ measured for the sample in which we could successfully identify spectral features.

Breaking the contamination level into the different selections, we find that the LBG\_PRI1, LBG\_PRI2, LBG\_PRI3 and LBG\_DROP samples have contamination rates of 32.5\%, 35.6\%, 38.2\% and 40.3\% respectively. Based on these recovered levels of contamination (and making the simplifying assumption that this applies to the faint unidentified spectroscopic sample), gives an average sky-density across our fields of $\approx1.8$ arcmin$^{-2}$ for all samples and $\approx1.3$ arcmin$^{-2}$ excluding the LBG\_PRI3 sample. Based on the volumes probed and the redshift distribution, these sky densities correspond to sky densities of $\sim4.0~h^3$Mpc$^{-3}$. 

\subsection{Galaxy redshifts}

In star-forming galaxies such as those presented here, the observed \lya\ emission is redshifted relative to the intrinsic galaxy redshift, while the interstellar absorption lines are blue-shifted \citep[see e.g. ][]{shapley03}. In \citetalias{2011MNRAS.414....2B}, we used the transformations given by \citet{adelberger05} in order to correct from the redshifts of the UV features to the intrinsic galaxy redshifts. These have now been superseded by those determined in \citet{2010ApJ...717..289S}, which we use in this paper and also apply to our previous data from \citetalias{2011MNRAS.414....2B}.

In \citetalias{2011MNRAS.414....2B}, we estimated the errors on the LBG redshifts using simulated spectra. Here, we extend the investigation into the redshift errors in our survey by using duplicate redshift measurements. The fields presented here, particularly Q2348, were designed with overlapping regions and consequently there are some LBG candidates which were observed in more than one mask. In cases where these duplicated targets are confirmed as LBGs, this provides two independent redshift measurements for the same LBG, and thus a direct observational test of the redshift measurement accuracy.

\begin{figure}
\centering
\includegraphics[width=8.0cm]{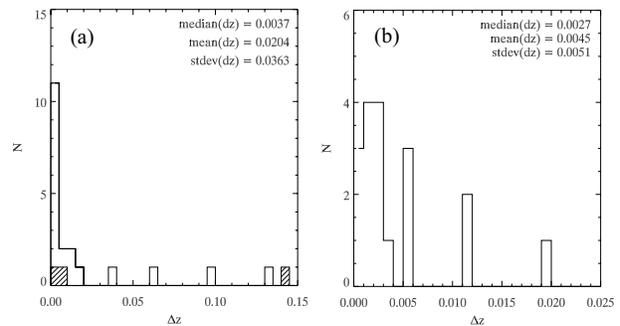}
\caption[Redshift measurement errors]{The distribution of redshift measurement errors, $\Delta z$, calculated using the LBGs which were observed twice in our survey and therefore have two independent redshift measurements. In panel \emph{(a)} we show the full distribution, in panel \emph{(b)} a close-up of the distribution at $z<0.025$. Overplotted as a hashed histogram in panel \emph{(a)} are the $\Delta z$ values for 3 sources in our survey which had a redshift in the survey of \citet{steidel03}.}
\label{f-dz}
\end{figure}

Fig.\ \ref{f-dz} shows the $\Delta z$ distribution for the LBGs with duplicate observations, where $\Delta z=|z_1-z_2|$ is the difference between the two redshift measurements. A total of 20 objects were classified as LBGs in two separate observations; of these, Fig.\ \ref{f-dz} indicates that 16 had fairly small errors of $\Delta z < 0.02$ (of which 13 had very small errors of $\Delta z < 0.005$), while 4 had considerably larger errors. In addition to these 20 objects, we have also searched a region of our Q0301 field which overlaps with \citet{steidel03} survey for any LBGs which were identified in both surveys: we find 3 such objects, and the redshift differences for these galaxies are also indicated in Fig.\ \ref{f-dz}.

The standard deviation of the 20 $\Delta z$ values measured for duplicate observations in our survey is $\sigma=0.036$, corresponding to a velocity error of $\approx$2,800 km \ps\ assuming a redshift of $z=2.8$, the sample mean.  However, this misrepresents the true error in our redshift measurements, since it is skewed by the 4 sources with very high $\Delta z$. These 4 values do not represent redshift \emph{measurement} errors, rather catastrophic outliers. In the cases where we find large $\Delta z$ values, the error does not arise due to uncertainty in the peak wavelength, but in uncertainty over which spectral feature is actually \lya. In these cases, different spectral features have been identified as \lya, leading to large $\Delta z$. These are therefore better characterised as \emph{identification} errors, in that two different solutions have been reached in the two observations.

For the 16 duplicated targets shown in Fig.\ \ref{f-dz}\emph{b}, the
same feature has been identified as \lya\ and therefore the $\Delta z$
for these objects gives an indication of the measurement error. The
standard deviation for these objects is $\sigma=0.005$, corresponding to
$\Delta v\approx380$ km \ps.

The suggestion, therefore, is that $\approx$80\% of our LBGs have
redshift measurement errors of $\Delta v \le 400$ km \ps, while the
other 20\% may have larger errors. This problem, however,
disproportionately affects sources with an ID quality parameter $Q=0.5$:
of the four sources with large $\Delta z$, one was given a quality
factor of 0.5 for both redshift measurements, while the other 3 have one
measurement with $Q=0.5$ and another with $Q=0.6$; in the latter cases
the $Q=0.6$ measurement is fairly robust while the $Q=0.5$ measurement
is less reliable. Therefore, the LBGs which may suffer from large errors
can be excluded by removing the $Q=0.5$ LBGs from the sample.

\subsubsection{Composite spectra} 

We have calculated composite spectra using the $z>2$ VIMOS low-resolution galaxy data. The composite spectra were generated by averaging over the spectra after having corrected the spectra for the instrument response and having masked skylines. In addition, each individual spectrum is normalised by its median flux in the rest-frame wavelength range $1250\AA\leq\lambda\leq1450$ before being combined to form the composite.

\begin{figure}
\centering
\includegraphics[width=8.0cm]{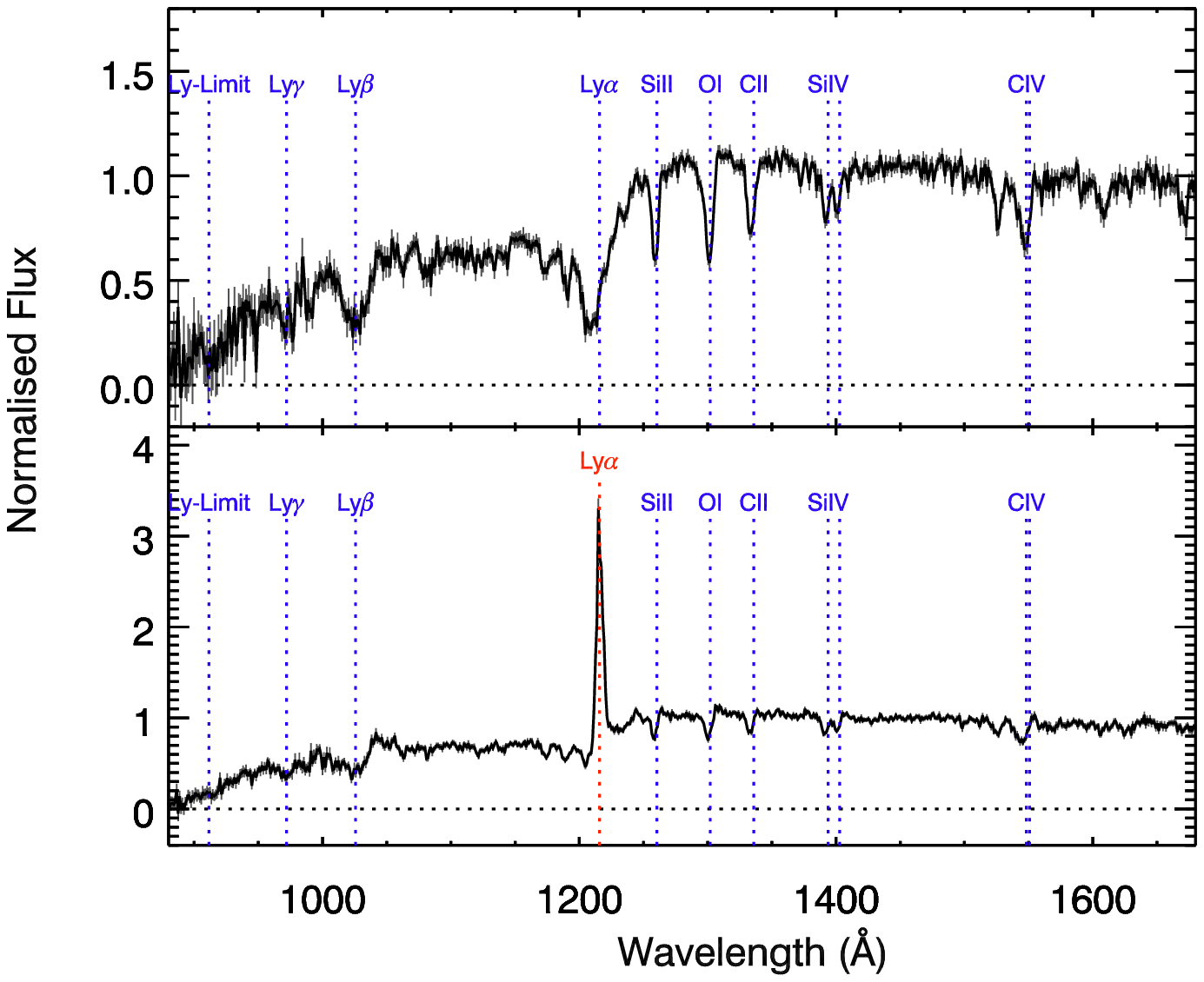}
\caption[Stacked spectra for LBGs showing \lya\ in emission]{Composite spectra for galaxies showing \lya\ in absorption (top) and emission (bottom), with ISM absorption lines clearly detected.}
\label{f-type-stack}
\end{figure}

In Fig.\ \ref{f-type-stack} we show stacked spectra for the LBGs, separated into those showing Ly$\alpha$ in emission (lower panel) and in absorption (upper panel). These stacked spectra show the average ultraviolet SED of a \zth\ LBG with excellent signal-to-noise, and the quality of these spectra provide an indication of the robustness of our LBG identifications.

\begin{figure}
\centering
\includegraphics[width=8.0cm]{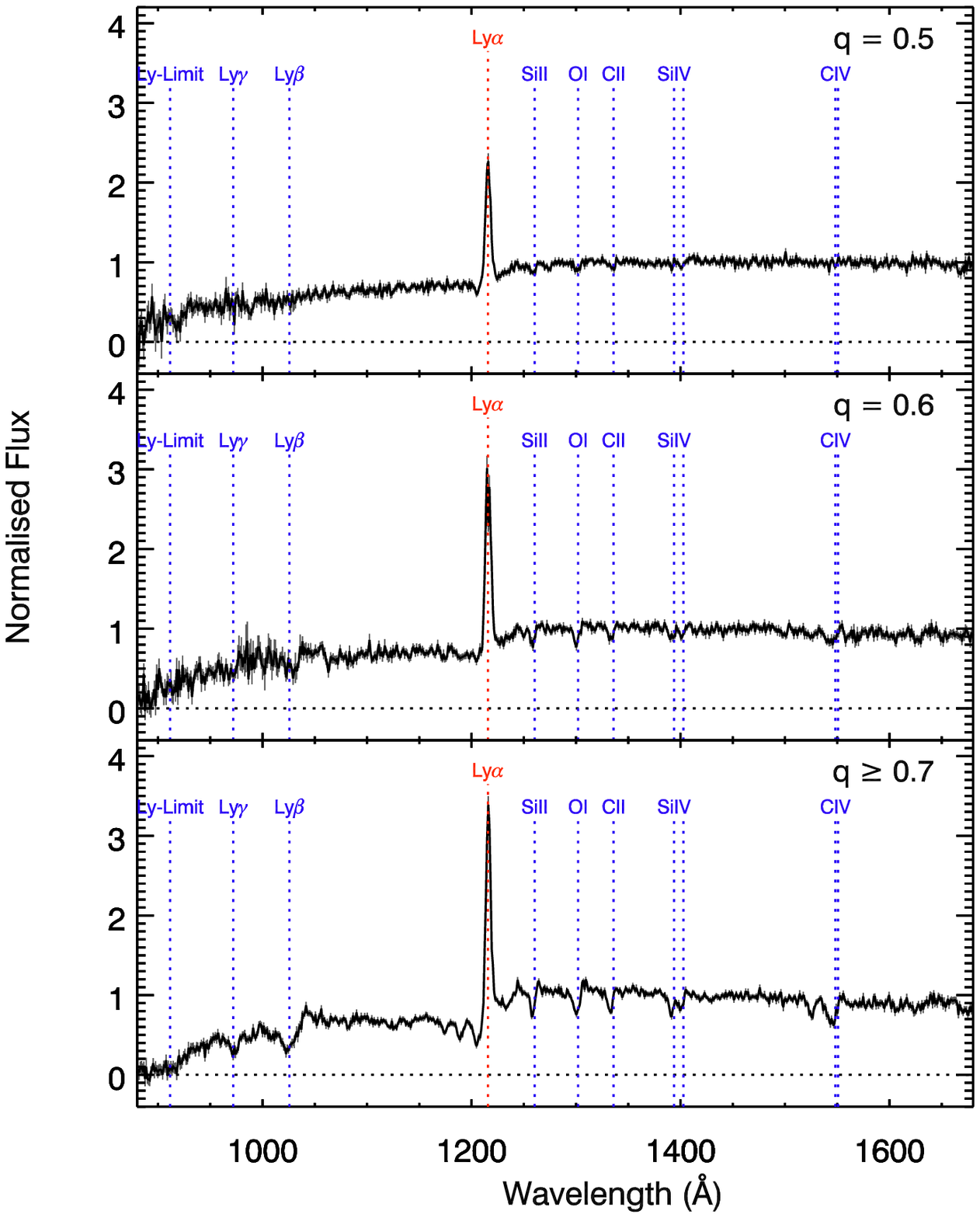}
\caption[Stacked spectra for LBGs with different ID quality parameters]{Composite spectra for galaxies classified as LBe's, separated by ID quality parameter $Q$. Those with $Q=0.5$ (upper panel) show comparatively weak \lya\ emission and marginal detections of ISM absorption lines. Moving to higher quality values, the strength of the \lya\ line increases and we also see high-SNR absorption features. Note that the spectra are not flux calibrated.}
\label{f-qstacks}
\end{figure}

In Fig.\ \ref{f-qstacks}, we show 3 separate composite spectra for sources classed as LBe's with quality IDs of $Q=0.5$, $Q=0.6$ and $Q\ge0.7$. These spectra reflect the quality criteria set out in \S\ref{s-targetid} well, with increasing quality spectra clearly showing increasingly high signal-to-noise in both \lya\ emission and ISM absorption features. In addition, the strength of the absorption features in the spectra appears to be systematically weaker with lower $Q$. This is likely the result of the lower signal-to-noise of the lower $Q$ identifications.

Finally we note that some potential flux is observed at wavelengths below the Lyman-limit, however even after stacking, the signal is subject to significant noise. Further analysis on the escape fraction may be possible using this data, but is beyond the scope of this paper.

\subsubsection{Quasars \& AGN}

\begin{figure}
\centering
\includegraphics[width=8.5cm]{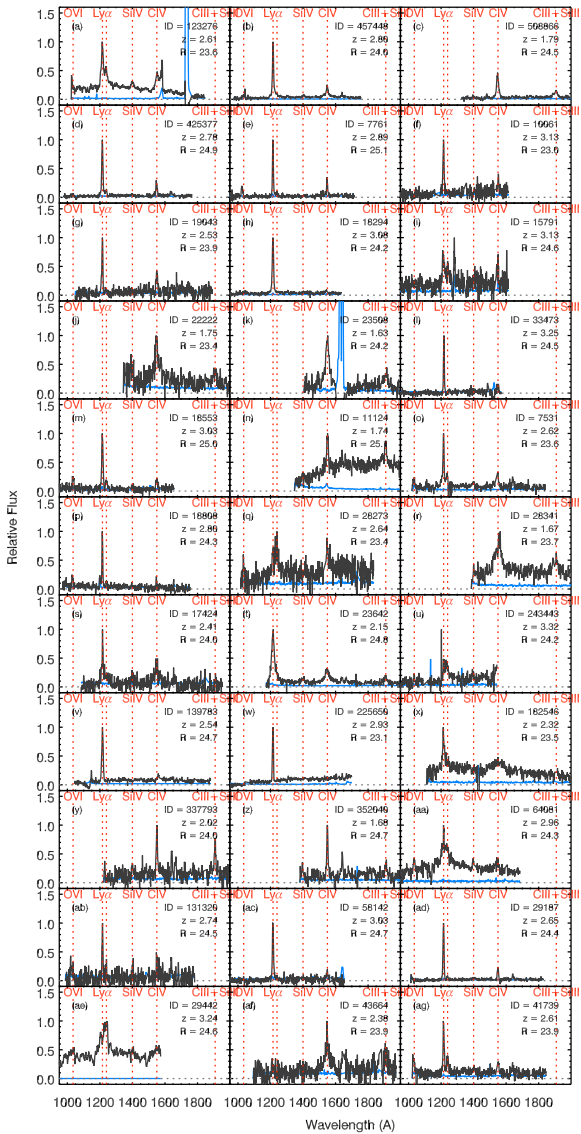}
\caption[Observed spectra for $z>2$ QSOs]{Rest-frame VIMOS spectra for the QSOs detected in the 5 fields presented. The $y$-axis scales differ from panel to panel, but the dotted line marks zero flux in each case. Dashed red lines indicate the wavelength of O\,\textsc{vi}, \lya, Ne\,\textsc{v} and C\,\textsc{iv} emission. Gaps in the spectra indicate that an artefact has been masked out.}
\label{f-qsospec}
\end{figure}

We have identified 33 $z>1.5$ AGN and QSOs in our spectroscopic sample, which we present here as part of the VLRS catalogue. They have been identified by the presence of strong Ly$\alpha$, C\,\textsc{iv} and C\,\textsc{iii}$+$Si\,\textsc{iii} emission lines as well as the generally weaker lines of O\,\textsc{vi}, N\,\textsc{v} and Si\,\textsc{iv}. Their spectra are shown in Fig.\ \ref{f-qsospec}. The above emission lines are indicated in each panel of Fig.\ \ref{f-qsospec}.

Several other emission lines are detected in some of the QSO spectra but are not marked in the figure. Ly$\beta$ $\lambda$1026 is clearly seen in panel (e), where it may be asymmetrically broadened to longer wavelengths by the presence of relatively weak O\,\textsc{vi} $\lambda$1035. Panel (p) shows an emission line peaking at 1029\AA, likely suggesting a blend between Ly$\beta$ and O\,\textsc{vi}.

Emission arising from the combination of O\,\textsc{i} $\lambda$1302 and Si\,\textsc{ii} 1304 is visible in a number of the spectra, for example in panels (a) and (h). Finally, many of the panels show clear emission at $\approx1400$\AA, arising from a blend of the Si\,\textsc{iv} $\lambda$1396 and O\,\textsc{vi}] $\lambda$1402 transitions.

The spectra show a clear mix of both broad and narrow line AGN, the narrow line objects suggestive of the presence of obscured AGN activity. These are reminiscent of the AGN identified in similar $z\sim3$ star-forming galaxy surveys, for example \citet{2002ApJ...576..653S,2011ApJ...733...31H,2012arXiv1206.3308H}.

\begin{figure}
\centering
\includegraphics[width=8.5cm]{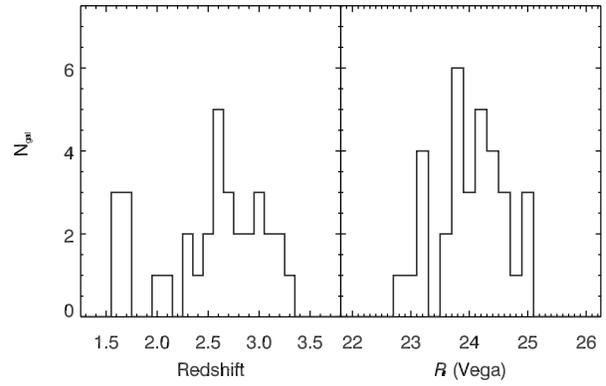}
\caption[]{The redshift (\emph{left}) and $R_{Vega}$ magnitude (\emph{right}) distributions of the 33 serendipitous spectroscopically identified quasar/AGN sample.}
\label{fig:qsoprops}
\end{figure}

The redshift distribution of the 33 $z>1.5$ AGN is shown in the left hand panel Fig.~\ref{fig:qsoprops}, whilst the $R$-band Vega magnitude distribution is shown in the right hand panel.

\section{Clustering of \zth\ LBGs}

We now analyse the clustering of the $z\approx3$ LBGs. As well as
offering insights into the growth and evolution of structure in the
Universe, we aim to measure the dynamics of the $z\approx3$ galaxy
population, i.e. peculiar velocities and gravitational infall, to inform 
the analysis of the gas-galaxy relationship via LBG-Ly-$\alpha$ forest 
cross-correlation (see Paper II).

We note that for the purposes of the clustering analysis we use the \numlbg\ $R\leq25$ VLRS sample (and place a limit of $R\leq25.2$ on the Keck sample with which it is compared and combined). Aside from this magnitude cut, all galaxies with $Q\geq0.5$ are included throughout this analysis. Taking the ${\cal R}=25.2$ limit for the Keck sample provides 815 galaxy redshifts within the \citet{steidel03} fields to combine with our VLRS sample.

In the analysis that follows, we measure the galaxy clustering as a function of galaxy-galaxy separation using the Landy-Szalay estimator:

\begin{equation}
\label{eq-xis}
\xi(x)=\frac{\left<\mathrm{DD}(x)\right>-2\left<\mathrm{DR}(x)\right>+\left<\mathrm{RR}(x)\right>}{\left<\mathrm{RR}(x)\right>}
\end{equation}

\noindent where $\xi(x)$ is the clustering as a function of separation $x$, $\mathrm{DD}(x)$ is the number of galaxy-galaxy pairs at that separation, $\mathrm{DR}(x)$ is the number of galaxy-random pairs and $\mathrm{RR}(x)$ is the number of random-random pairs. This is estimated using a random catalogue that consists of 20$\times$ as many random points as data points and that covers an identical area. The redshift distribution of the random catalogue is set using a polynomial fit to the data.

We focus on fitting the auto-correlation function in the semi-projected, $w_p(\sigma)$, and full 2-D, $\xi(\sigma,\pi)$, forms, where $\sigma$ and $\pi$ are  the separation of two galaxies transverse and parallel to the line-of-sight. But we shall also study the LBG $z$-space correlation function, $\xi(s)$, where the signal can be significantly higher at large scales.

\begin{figure}
\centering
\includegraphics[width=\columnwidth]{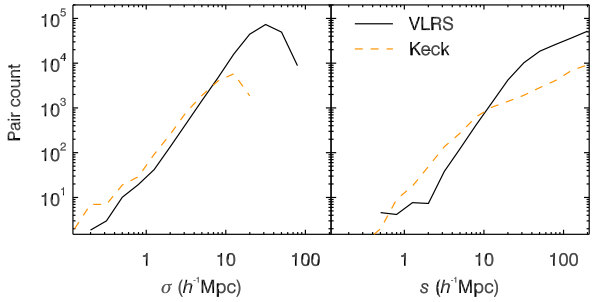}
\caption{Pair counts (DDs) as a function of pair separation in the
transverse, $\sigma$, and 3D Hubble, $s$, directions. The Keck data
(dashed orange curves) provide better sampling at small scales, whilst
the VLRS data (solid black curves) provides sampling at larger scales
that is not provided by previous data.}
\label{fig:pairs}
\end{figure}

For  $\xi(\sigma,\pi)$ in particular, we also consider the combined sample of the VLRS data with the Keck LBG data of \citet{steidel03}. The Keck data offers higher sampling rates than the VLRS, but across smaller field sizes ($\approx8$\arcmin). This is illustrated in Fig.~\ref{fig:pairs} where the solid black line shows the VLRS pair counts (DD) as a function of separation in the transverse direction, $\sigma$ (left hand panel), and the 3-D separation, $s=\sqrt{\sigma^2+\pi^2}$ (right hand panel). In both panels the Keck pair counts are shown by the dashed orange line. Fig~\ref{fig:pairs} shows that the VLRS pair counts in both the transverse  and  3-D distance are significantly higher than in the previous Keck sample at $\sigma,s\ga10$h$^{-1}$Mpc.

On the validity of combining our own LBG sample with that of \citet{steidel03}, we note that \citet{steidel03} used photometry with mean 1$\sigma$ depths of $\left<\sigma(U_n)\right>=28.3$, $\left<\sigma(G)\right>=28.6$ and $\left<\sigma({\cal R})\right>=28.0$, whilst their imposed ${\cal R}$ band limit was ${\cal R}=25.5$. Using the transformations of \citet{1993AJ....105.2017S}, the \citet{steidel03} 1$\sigma$ limits correspond to $U=27.55$, $B=28.77$ and $R=27.86$ in the Vega system. Comparing this to the average depths ascross all the VLRS fields, we have mean $3\sigma$ depths of $U=25.8$, $B=26.4$ and $R=25.9$, which equate to $1\sigma$ depths of $U=27.0$, $B=27.6$ and $R=27.1$, approximately 1 mag fainter in each band than the \citet{steidel03} imaging data. However, given our limit of $R=25$ (and our imposed ${\cal R}=25.2$ for the Keck sample), the $B-R$ (and $G-{\cal R}$) constraints on the selections, and the inclusion of galaxies with no $U$-detection, the VLRS and Keck samples will be relatively equivalent in terms of the galaxies included.

It is clear however, that the VLRS sample, although giving a close approximation of the Keck sample selections, is not a perfect replica of the Keck selection. Given the difference in the filters and the moderate difference in depths this was unlikely to ever be the case. The redshift distributions are relatively well matched, but (partially due to the addition of the LBG\_PRI3 selection) the VLRS sample is skewed somewhat to marginally lower redshifts (as illustrated in Fig.~\ref{}). Additionally, the sky and space densities are close but not perfectly matched, as are the $R-band$ magnitude distributions \citepalias[as shown by][]{2011MNRAS.414....2B}. As a result, the UV luminosity functions will be similarly close but not perfectly matched. In combining the two samples we therefore note these differences and use the results of combining the samples with caution. However, it is beneficial to do so in order to help constrain the redshift-space distortion effects, which are an important element of further work incorporating the Ly$\alpha$ forest to constrain the distribution of gas around these star-forming galaxies. Furthermore, it is difficult to perform these tests with either sample alone given the Keck sample's small area coverage and the VLRS sample's comparatively lower sampling rate. Therefore, although the combination is not ideal, it offers an indication of the impact of redshift space distortions on the correlation functions that may be utilised in subsequent work.

\subsection{Semi-projected correlation function, $w_p(\sigma)$}

We first estimate the LBG clustering using the semi-projected correlation
function, $w_p(\sigma)$. This gives the clustering at fixed transverse separation, $\sigma$, integrated
over line-of-sight distance, $\pi$, approximately independent of
the effect of peculiar velocities and is given by:

\begin{equation}
\label{eq:wpsig}
w_p(\sigma) = 2\int^{\infty}_0\xi(\sigma,\pi)d\pi
\end{equation}

\noindent where $\xi(\sigma,\pi)$ is the 2-D auto-correlation function.
We integrate $\xi(\sigma,\pi)$ over the range
$0<\pi<\pi_\mathrm{max}(\sigma)$, where $\pi_\mathrm{max}(\sigma)$ is
given by the maximum of $1000(1+z)/H(z)$ and $15\sigma$ at a given sky
separation $\sigma$ (consistent with
\citealt{adelberger03,daangela05b}).

In the calculation of $w_p(\sigma)$, we make a correction for the effect
of `slit collisions', following \citetalias{2011MNRAS.414....2B}. Any
object observed with VIMOS takes up an area on the detector of at least
$40\times640$ pixels (\S\ref{ss-specobs}), corresponding to
$8\marcsec\times130\marcsec$ on-sky. Other candidates lying within this
area can therefore not be observed (unless the area is revisited), and
as a result, pairs of LBGs at small separations are systematically
missed by our survey. This effect will reduce the measured LBG
auto-correlation at small separations. \citetalias{2011MNRAS.414....2B}
quantified this effect by comparing the angular auto-correlation function
of photometrically selected LBG candidates and spectroscopically
observed candidates. Using their result, we correct for this effect in
our LBG survey by weighting DD pairs at $\theta<8\marcmin$ according to
the weighting factor given by

\begin{equation}
\label{eq-slitcollisions}
W(\theta)=\frac{1}{1-(0.0738\times\theta^{-1.052})}
\end{equation}

\noindent where $\theta$ is the angular separation in arcminutes.

In addition to the slit collision correction, a further correction - the integral constraint - is required to compensate for the effect of the limited field sizes. For this we follow the commonly used approach of using the random-random pair distributions, which have been constructed to match the survey geometry, to determine the magnitude of the integral constraint. This method has been well described by a number of authors \citep[e.g.][]{1977ApJ...217..385G,1980lssu.book.....P,1993MNRAS.263..360R,1996MNRAS.283L..15B,2006A&A...457..145P}, with \citet{2006A&A...457..145P} in particular providing a detailed discussion in relation to the projected correlation function, and we provide a brief description of the calculation here.

The measured correlation function is given by the true correlation function minus the integral constraint, ${\cal I}$:

\begin{equation}
w_m(\sigma)=w_t(\sigma) - {\cal I}
\end{equation}

Assuming a power-law form for the the real-space clustering, the true projected clustering is fit by:

\begin{equation}
w_{p}(\sigma) =  Cr_{0}^{\gamma}\sigma^{1-\gamma}
\end{equation} 

\noindent where $r_0$ is the real-space clustering length and $\gamma$ is the slope of the real-space clustering function, $\xi(r)$, which is characterised by a power-law of the form:

\begin{equation}
\xi(r)=\left(\frac{r_0}{r}\right)^\gamma
\end{equation}

The factor $C$ is given by:
 
\begin{equation}
C = \left(\frac{\Gamma\left(\frac{1}{2}\right)\Gamma\left(\frac{\gamma-1}{2}\right)}{\Gamma\left(\frac{\gamma}{2}\right)}\right)
\end{equation}

\noindent where $\Gamma$ is the Gamma function. Given this framework, the integral constraint can be estimated from the mean of the random-random pair counts, $\left<RR\right>$, and the slope of the correlation function, such that:

\begin{equation}
\frac{{\cal I}}{Cr_0^\gamma} = \frac{\Sigma\left<RR(\sigma)\right>\sigma^{1-\gamma}}{\Sigma\left<RR(\sigma)\right>}
\end{equation}

\begin{figure}
\centering
\includegraphics[width=\columnwidth]{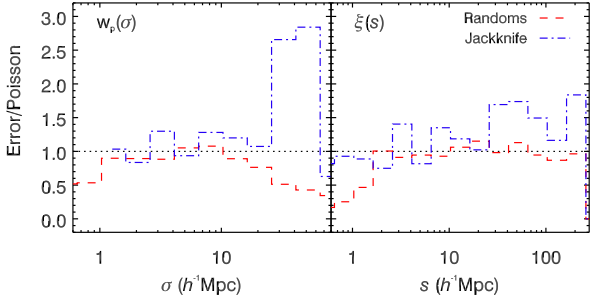}
\caption{Ratio of the random mock (red dashed histogram) and the jack-knife (blue dot-dashed histogram) error estimates to the Poisson errors as a function of separation for the projected correlation function (left panel) and the redshift space autocorrelation function (right panel). The results are shown for the VLRS sample only.}
\label{fig:xis_err}
\end{figure}

Quantifying errors on the auto-correlation function has been performed using Poissonian, jack-knife and random realisation error estimates. The Poisson errors are given by:

\begin{equation}
\Delta\xi = \frac{(1+\xi)}{\sqrt{\left<DD\right>/2}}
\end{equation}

The jack-knife errors were computed by splitting the data into individual fields, with the large fields (i.e. HE0940 and Q2348) being split into two fields each. We therefore have 11 different jack-knife realisations with a single field (or half-field) being excluded in each realisation.

The random realisation error estimates incorporate 100 random catalogues with the same number of objects as the real data. We then calculate the correlation function using these random realisations to calculate the $\left<DD\right>$ pairs and take the standard deviation of the results as the uncertainty on the measurement.

In Fig.~\ref{fig:xis_err}, we compare the above error estimates for
$w_p(\sigma)$ and the redshift space clustering function, $\xi(s)$ (see
section~\ref{sec:xisp}), showing the ratio of the jack-knife and random
realisation methods to the Poisson result. The estimates are consistent
with each other over scales from $\approx1-25\hmpc$. In
what follows, we therefore use the Poisson estimates at separations of
$<25\hmpc$ and jack-knife estimates at separations 
$>25\hmpc$.

\begin{figure}
\centering
\includegraphics[width=\columnwidth]{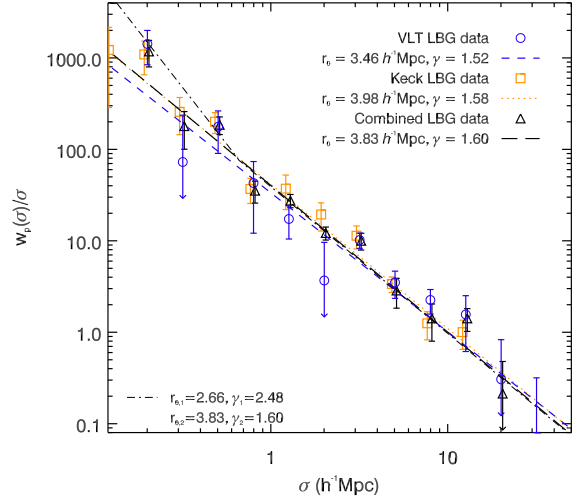}
\caption{Projected auto-correlation function, $w_p(\sigma)/\sigma$, for the VLT LBG sample (black circles). The result for the $2.5<z<3.5$ Keck LBG sample is shown by the orange squares and error bars for both results are estimated using Poisson errors. The best fit power laws for both the VLT (solid black line) and Keck (dashed orange line) are shown, with clustering amplitudes of $r_0=\vwprn\pm\vwprne\hmpc$ and $r_0=\kwprn\pm\kwprne\hmpc$, with slopes of $\gamma=\vwpgam\pm\vwpgame$ and $\gamma=\kwpgam\pm\kwpgame$ respectively. The cyan triangles show the combined result and the dotted cyan line the best fit to this of $r_0=\cwprn\pm\cwprne\hmpc$ with a slope of $\gamma=\cwpgam\pm\cwpgame$. The black dash-dot line shows a double power-law, motivated by \citetalias{2011MNRAS.414....2B}, with $r_0=2.61~\hmpc$ and $\gamma=2.48$ below the break and $r_0=3.75~\hmpc$ and $\gamma=1.61$ above it.}
\label{fig:wpsig}
\end{figure}

The projected auto-correlation function for the VLRS sample (black
circles), the Keck sample \citep[orange squares; ][]{steidel03} and the
two combined (cyan triangles) is shown in Fig~\ref{fig:wpsig}. The
plotted data include the integral constraint correction, which we
estimated as ${\cal I}_{wp}=\vwpic$ and ${\cal I}_{wp}=\kwpic$ for the VLRS and Keck data respectively.

Based on this we estimate a clustering length of the entire VLRS sample
of $r_0=\vwprn\pm\vwprne\hmpc$ (comoving) with a slope of
$\gamma=\vwpgam\pm\vwpgame$. The Keck result on it's own gives a result
of $r_0=\kwprn\pm\kwprne\hmpc$ with $\gamma=\kwpgam\pm\kwpgame$, whilst
the combined VLRS$+$Keck data gives $r_0=\cwprn\pm\cwprne\hmpc$ with a
slope of $\gamma=\cwpgam\pm\cwpgame$. These $r_0$ results are comparable
to the clustering of star-forming galaxies at lower redshifts
\citep[e.g.][]{2009MNRAS.395..240B,2010MNRAS.403.1261B}.

Comparing to other measurements of the $z\approx3$ LBG clustering length,
\citet{2001ApJ...550..177G} measured $r_0=5.0\pm0.7\hmpc$ for
$R_{AB}<25$ LBGs,  but for  a relatively small number of galaxies
($\approx400$). Building on that sample, \citet{adelberger03} measured
$r_0=3.96\pm0.29\hmpc$ with a slope of $\gamma=1.55\pm0.15$ at
$R_{AB}<25.5$. We note that with the same sample, but a different
method, \citet{2005ApJ...619..697A} found a higher clustering strength
of $r_0=4.5\pm0.6\hmpc$. Subsequently, \citet{2006ApJ...652..994C}
measured the clustering of $z\approx3$ LBGs in fields around damped
Ly$\alpha$ absorbers and found a lower clustering strength of
$r_0=2.65\pm0.48\hmpc$ with a slope of $\gamma=1.55\pm0.40$ at
$R_{AB}<25$, whilst \citet{2012ApJ...752...39T} performed a similar measurement but around $z\sim2.7$ QSOs (and with a galaxy sample incorporating a mixture of LBG and BX selections) and found a clustering length of $r_0=6.0\pm0.5~\hmpc$. Overall, our result appears consistent with other measurements, although marginally lower than the \citet{2001ApJ...550..177G,adelberger03,2005ApJ...619..697A} results, which are all based on the same - or a subset of the same - sample. As observed by some of the above authors, the LBG clustering lengths are generally somewhat smaller than those measured for the slightly lower-redshift BM and BX selections.

The above estimates are based on spectroscopically confirmed samples and
a number of clustering measurements exist based on photometric samples.
For example \citet{foucaud03} measured $r_0=5.9\pm0.5\hmpc$ from the
angular correlation function of $R_{AB}<24.5$ LBGs from the
Canada-France Deep Fields Survey \citep{2001A&A...376..756M}, a higher
$r_0$ than the spectroscopic samples, but also a significantly brighter
magnitude cut. Additionally, \citet{2005ApJ...619..697A} measured
$w(\theta)$ for photometrically selected LBGs and found $r_0=4.0\pm0.6\hmpc$ for $R\leq25.5$ LBGs, consistent with our results. \citet{hildebrandt07}  measured $r_0=4.8\pm0.3\hmpc$ for $22.5<R_{\mathrm{Vega}}<25.5$ galaxies in the GaBoDS data. Subsequently, \citet{2009A&A...498..725H} measured  $r_0=4.25\pm0.13\hmpc$ for CFHTLS
LBGs at $r_{AB}<25$ and using photo-z from {\small HYPERZ} \citep{2000A&A...363..476B}. In general, the clustering measured from photometric samples appears to give somewhat larger clustering lengths than those obtained for the spectroscopic samples. As with our own sample however, these selections are not perfect replicas of the original $U_nG\cal R$ based selection and this may be part of the cause of this, perhaps resulting in subtle differences in the redshift or luminosity ranges.

\subsection{2D Auto-Correlation Function, $\xi(\sigma,\pi)$}
\label{sec:xisp}

As discussed above, integrating along the redshift/line-of-sight
direction leaves the semi-projected correlation function, $w_p(\sigma)$,
independent of the effects of galaxy peculiar motions. We now attempt to
fit the full 2D correlation function, $\xi(\sigma,\pi)$, to retrieve the
kinematics of the galaxy population and to make new estimates of $r_0$.

\begin{figure}
\centering
\includegraphics[height=0.25\textwidth]{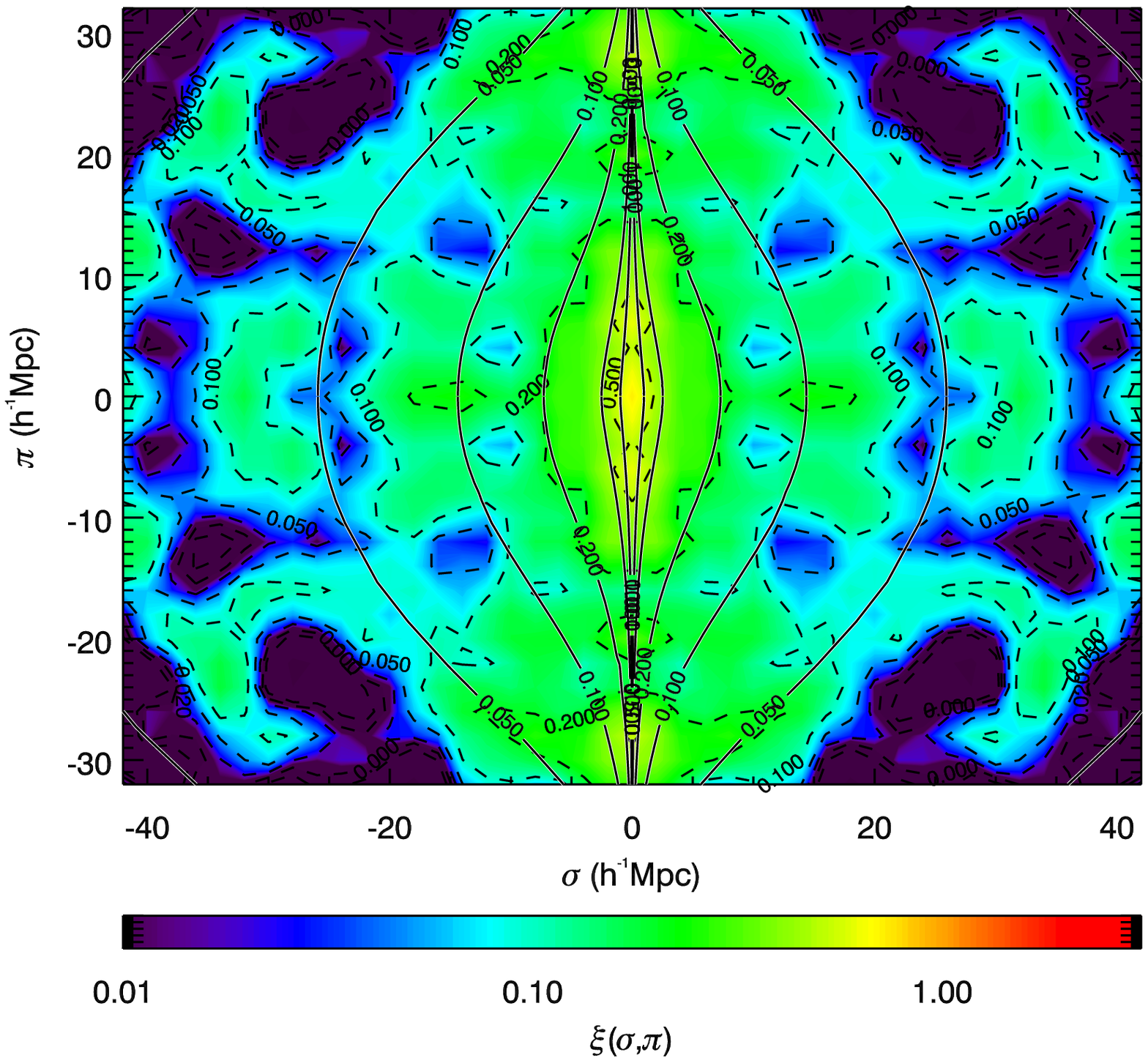}
\includegraphics[height=0.25\textwidth]{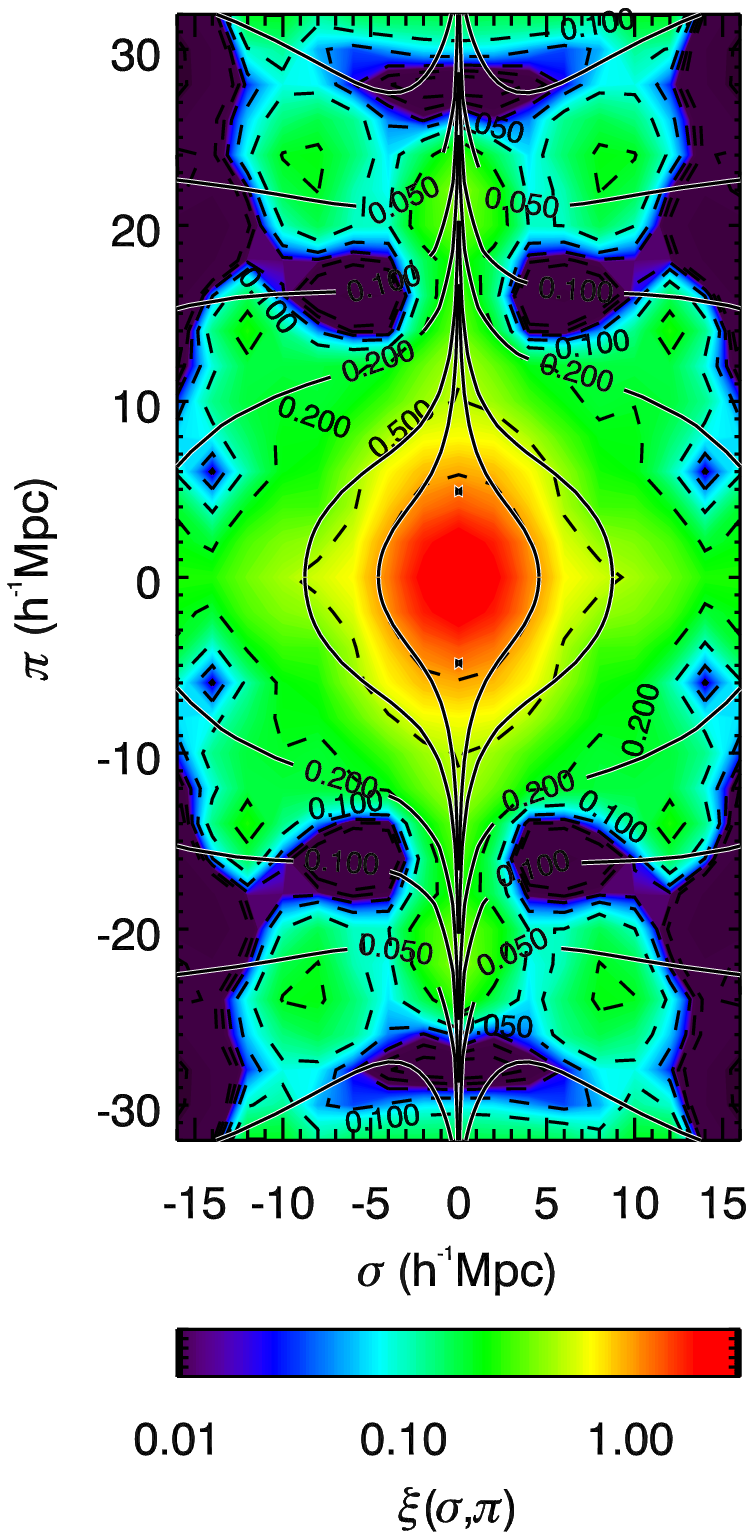}
\caption{The two-dimensional auto-correlation function,
$\xi(\sigma,\pi)$ results for the VLRS (left) and Keck (right) data
samples individually. The shaded contour map and dashed contour lines
show the measured $\xi(\sigma,\pi)$ in each case.}
\label{fig:xisp}
\end{figure}

As before, we use the Landy-Szalay estimator to calculate the
correlation function but now as a function of both transverse
separation, $\sigma$, and line-of-sight separation, $\pi$. We use the
same random catalogues matching the survey fields as used for the
calculation of the projected correlation function. We again calculate
the integral constraint for the data sets using the random catalogues via:

\begin{equation}
\frac{{\cal I}}{r_0^\gamma} = \frac{\Sigma\left<RR(s)\right>s^{\gamma}}{\Sigma\left<RR(s)\right>}
\end{equation}

\noindent where $s=\sqrt(\sigma^2+\pi^2)$. This gives values of ${\cal
I}_\xi=\vxiic$ and ${\cal I}_\xi=\kxiic$ for the VLRS and Keck data
samples respectively.

\begin{figure*}
\centering
\includegraphics[height=0.4\textwidth]{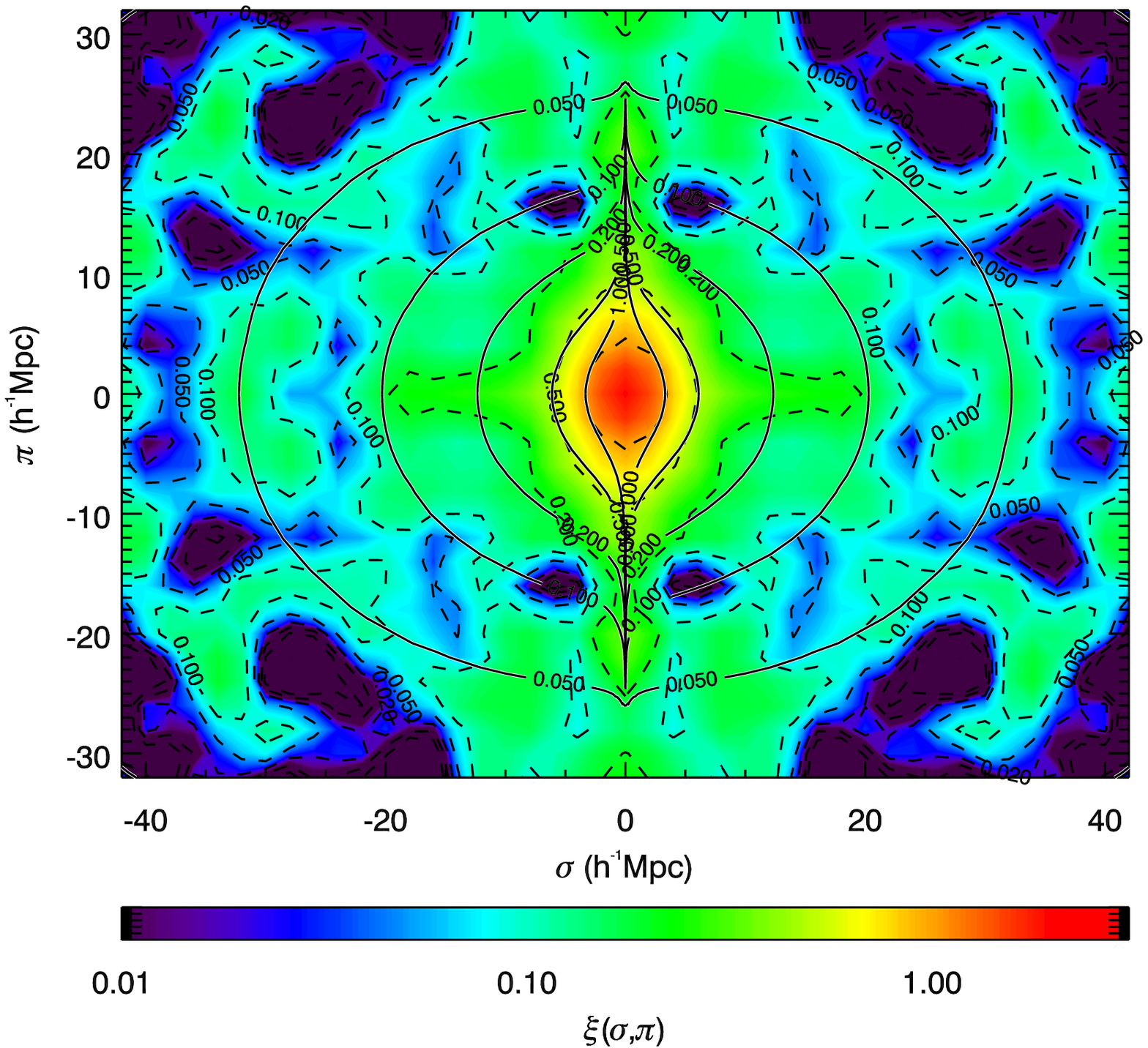}
\includegraphics[height=0.4\textwidth]{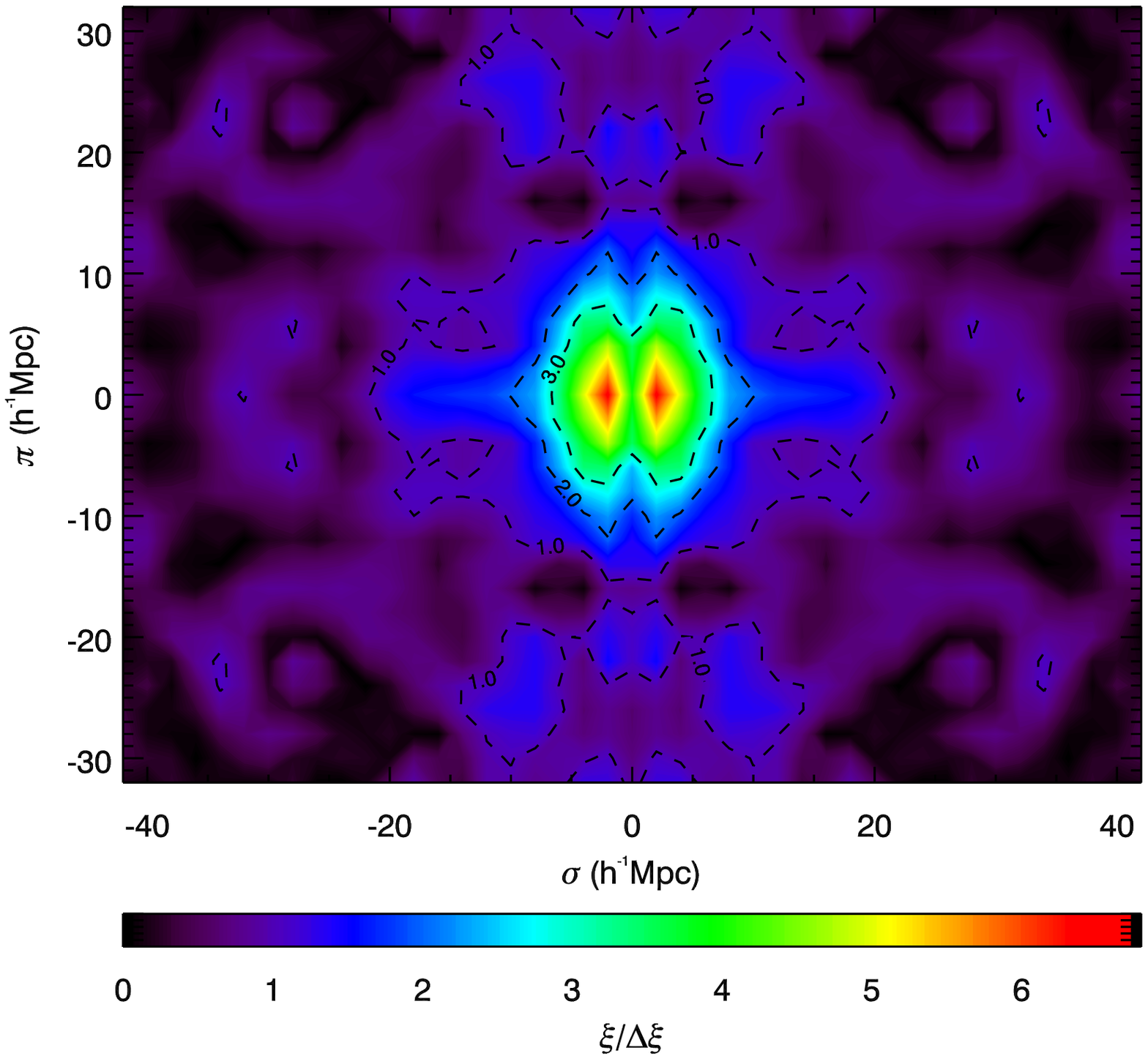}
\caption{The two-dimensional auto-correlation function, $\xi(\sigma,\pi)$ results for the VLRS sample (left) and the signal-to-noise on the result (right). The shaded contour map and dashed contour lines show the measured $\xi(\sigma,\pi)$, whilst the solid contour lines give the best fitting model.}
\label{fig:xisperr}
\end{figure*}

Fig.~\ref{fig:xisp} shows the result for the VLRS data (left-hand
panel), which provides a greater handle on the large scale ($s\gtrsim10
h^{-1}$Mpc) clustering, the Keck data (right panel), which provides
greater sampling on small scales. Fig. \ref{fig:xisperr} shows the VLRS
and Keck results combined. In each case, $\xi(\sigma,\pi)$ was
calculated in linear $2\hmpc$ bins and subsequently smoothed with a fwhm of $2\hmpc$.

For both the VLRS and Keck samples, we see the `finger of god' effect at
small $\sigma$ scales in which the clustering power is extended in the
$\pi$ direction. This effect is a combination of galaxy peculiar
velocities and measurement errors on the galaxy redshifts. In addition,
in the VLRS a flattening of the clustering measurement at large scales
is evident, which is caused by dynamical infall of galaxies. 

We now fit models of the clustering to these results, initially assuming
a single power-law for $\xi(r)$ and  allowing $r_0$ and the kinematical 
parameters to vary. We take the  $r_0$ and $\gamma$ estimates 
from the $w_p(\sigma)$ fit as the starting point in fitting the
2D clustering. The kinematics are characterised by two parameters: the
velocity dispersion in the line of sight direction
$\sqrt{\left<w_z^2\right>}$ and the infall parameter, $\beta$. The
model we use incorporating the galaxy kinematics is described in full by
\citet{hawkins03} and \citetalias{2011MNRAS.414....2B}. The model accounts for two key affects on the clustering statistics caused by galaxy motions. The first is the finger-of-god effect, which is constrained by the velocity dispersion and the second is the Kaiser effect (i.e. the coherent motion of galaxies on large scales), which is characterised by $\beta$.

For the VLRS and the combined samples we fit over the range $1.0\leq s\leq25\hmpc$, whilst for the Keck data by itself we limit the fit to the scales $1.0\leq s\leq15\hmpc$ (note that the largest single field available in the Keck data is $\approx15\hmpc$).

For the two samples individually, we find that it is difficult to place
reasonable constraints on both the velocity dispersion and the infall
together. With the VLRS data (over the range $1\leq(\sigma,\pi)\leq25\hmpc$), we find $\beta(z=3)=0.3^{+1.7}_{-0.3}$ and
$\sqrt{\left<w_z^2\right>}=1700^{+2000}_{-900}\kps$, the low
signal-to-noise on small scales limiting the fit accuracy. We
experimented with adding a uniform error distribution out to
$\pm12000\kps$ to the Gaussian velocity dispersion (c.f. Fig. \ref{f-dz})
but this made little difference in the $\sigma,\pi$ range fitted.
Fitting the Keck data gives best fit values of $\beta(z=3)=0.85^{+0.30}_{-0.35}$ and $\sqrt{\left<w_z^2\right>}=700\pm220\kps$. We note that \citet{daangela05b} performed a similar fit to the Keck data for
$\beta(z=3)$, but kept a constant velocity dispersion of
$\sqrt{\left<w_z^2\right>}=400\kps$, finding a value for the infall
parameter of $\beta(z=3)=0.15^{+0.20}_{-0.15}$. By also setting the
velocity dispersion to a value of $400\kps$, we find that we retrieve a
comparable result to \citet{daangela05b}, highlighting the degeneracy
between the velocity dispersion and the infall parameter.

\begin{figure}
\centering
\includegraphics[width=8.0cm]{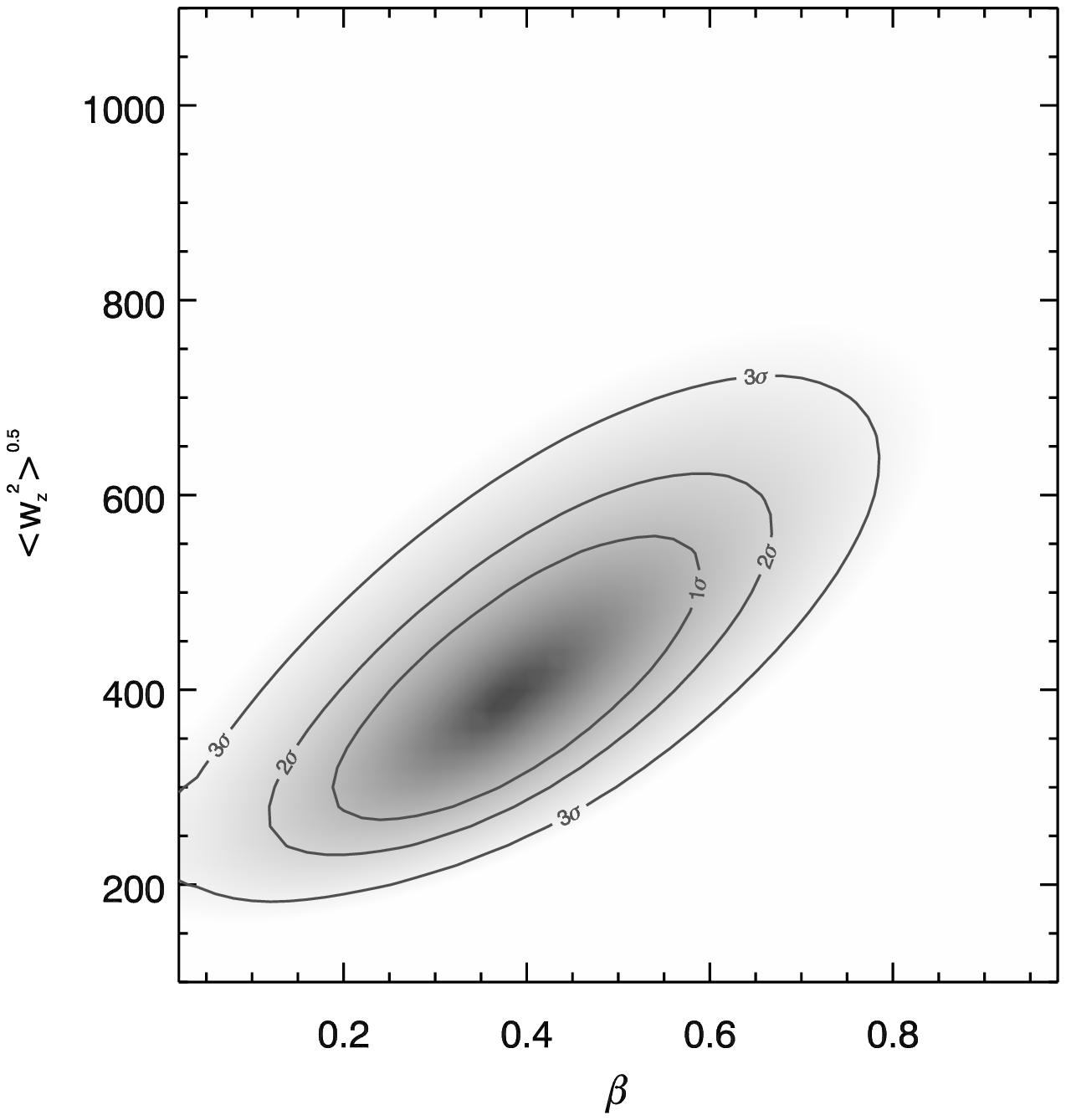}
\caption[Redshift measurement errors]{Fitting contours for peculiar
velocity and bulk inflow based on the combined VLRS$+$Keck
$\xi(\sigma,\pi)$. The best fitting result is given by
$\beta(z=3)=\infpar\pm\infpare$ and
$\vdisp=\vdispar^{\vdisparup}_{\vdispardown}\kps$.}
\label{fig:vdisp_beta_fit}
\end{figure}

Ultimately, fitting the VLRS $\xi(\sigma,\pi)$ is hindered by a lack of
signal-to-noise on small scales, whilst the fit to the Keck data is
hindered by the small size of the fields. We thus combine the two
datasets and fit the full LBG sample in the same manner as with the
individual samples. The fit is performed in the range $1<s<25\hmpc$ and
we allow the velocity dispersion and the infall parameter to vary. The
resulting fit gives a velocity dispersion of
$\vdisp=\vdispar^{\vdisparup}_{\vdispardown}\kps$ and an infall
parameter of $\beta(z=3)=\infpar\pm\infpare$. We show the $\chi^2$
contours for the fit in the $\beta(z=3)-\sqrt{\left<w_z^2\right>}$ plane
in Fig.~\ref{fig:vdisp_beta_fit} (the contours represent the $1\sigma$,
$2\sigma$ and $3\sigma$ confidence limits). From this figure, the
degeneracy can be seen between $\sqrt{\left<w_z^2\right>}$ and $\beta$,
where increasing $\beta$ similarly increases the best fit velocity
dispersion. The best fitting results are plotted over the contour maps of the $\xi(\sigma,\pi)$ measurements in Fig.~\ref{fig:xisperr} (dashed contours). As with the data, we see the finger-of-god and large scale flattening effects in the fitted models.

\subsection{Redshift-space correlation function, $\xi(s)$.}
\label{sec:xis}

\begin{figure}
\centering
\includegraphics[width=8.0cm]{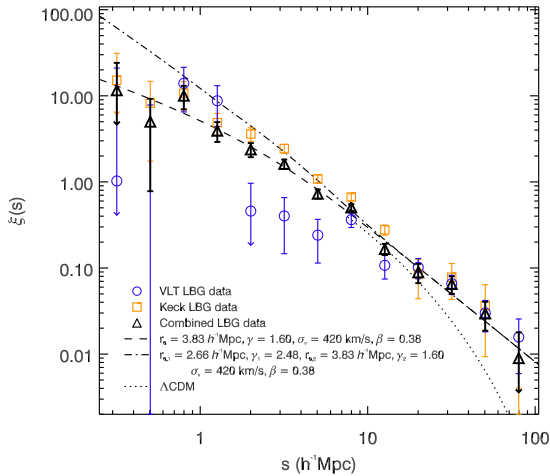}
\caption{Redshift space correlation function, $\xi(s)$, for the VLRS (open blue circles), Keck (orange squares) and combined (black triangles) samples. The short and long dashed curves show single power law model fits to the combined sample based on fits to $w_p(\sigma)$ and $\xi(\sigma,\pi)$. The dash-dot curve gives a double power law model fit, whilst the dotted curve at large scales shows the predicted $\Lambda$CDM clustering.}
\label{fig:xis}
\end{figure}

In order to check the consistency of our measurements, we now compare
the model fit obtained from $w_p(\sigma)$ and $\xi(\sigma,\pi)$ to the
measured redshift-space auto-correlation function $\xi(s)$. Again we use
the Landy-Szalay estimator and quote errors based on Poisson estimates.
The $\xi(s)$ results for the VLRS, Keck and combined LBG samples are shown in Fig.~\ref{fig:xis}. We also plot the single power law estimate of the intrinsic clustering from our fits to the VLT$+$Keck $\xi(\sigma,\pi)$ (dotted line) and the result of this power-law after applying the best fit values for $\beta$ and $\vdisp$ (dashed line). The final fit
incorporating the galaxy dynamics is marginally low compared to the
data-points, but is easily consistent within the error bars.

Our measurements of $\beta$ and $\vdisp$ are consistent with the previous measurements using the first VLT dataset \citepalias[][$\beta=0.48$ and $\vdisp=700~\kps$]{2011MNRAS.414....2B}. As discussed in \citetalias{2011MNRAS.414....2B}, the median measurement error on the galaxy lines on the VLT VIMOS spectra is $\approx350\kps$. In addition, an uncertainty of $\approx100\kps$ is introduced by the transformation from outflow redshifts to intrinsic galaxy redshifts \citep{2010ApJ...717..289S}. The final contribution to the velocity dispersion is from the intrinsic peculiar velocities of the galaxies. Using the GIMIC simulations \citep{2009MNRAS.399.1773C} we have analysed the mean velocity dispersion of LBG-like galaxies and find a value of $\approx100\kps$. Combining these three elements in quadrature, we would expect a pairwise velocity dispersion of $\sqrt{\left<w_z^2\right>}\approx\sqrt{2}\times\sqrt{(350\kps)^2+(100\kps)^2+(100\kps)^2}\approx500\kps$. This is within the $1\sigma$ error contours given in Fig. \ref{fig:vdisp_beta_fit}. This value is also reasonably consistent with the VLRS $\xi(s)$ estimate (see  Fig.~\ref{fig:xis}).

We also show in Fig.~\ref{fig:xis} the matter correlation, $\xi(r)$, scaled to the LBG clustering strength. This was calculated using the CAMB software and using a flat $\Lambda$CDM cosmology with $\Omega_m=0.27$, $H_0=70\kps$Mpc$^{-1}$ and
$\Omega_b h^2=0.022$. There are currently some claims that
non-Gaussianity is detected at $z\approx1$ in NRAO VLA Sky Survey radio
source \citep{2010ApJ...717L..17X} and Luminous Red Galaxy (LRG) datasets
\citep{2011PhRvL.106x1301T,2011MNRAS.416.3033S,2012arXiv1204.3609N}. The evidence generally comes comes from detecting large scale excess power via flatter slopes for angular correlation functions. Since non-Gaussianity is easier to detect at high redshift this motivates checking the  LBG $\xi(s)$ for an excess. We have already noted that the slope from $w_p$ and $\xi(\sigma,\pi)$ at $\gamma=1.55$ is much flatter than the canonical $\gamma=1.8$. This slope is also flatter than the $z\approx1$ LRG large-scale $w(\theta)$ slope of \citet{2012arXiv1204.3609N}. We see that the VLRS does give reasonably accurate measurements for $25<s<100\hmpc$ and that the observed LBG $\xi(s)$ shows a $\approx2.5\sigma$ excess over the $\Lambda$CDM model in this range. Even when the  marginally smaller integral constraint for the $\Lambda$CDM model is assumed the discrepancy remains at $\approx2\sigma$. We conclude that there is some evidence for an excess over the standard $\Lambda$CDM model but independent LBG data is needed to confirm this on the basis of the redshift space correlation function. The statistical error on the LBG $w(\theta)$ from \citetalias{2011MNRAS.414....2B} is smaller but the flat power-law here is only seen to $\theta=10'$ or $r=13-14\hmpc$ and this is not enough to decide the issue.

\subsection{Double power-law correlation function models}

We next look to see if a more complicated model than a power-law for $\xi(r)$ is required. This is motivated firstly because \citetalias{2011MNRAS.414....2B} noted that there was an increase in the slope at $\approx1\hmpc$ in the LBG angular auto-correlation function, $w(\theta)$, suggestive of the split between the 1-halo and 2-halo terms in the halo model of clustering. Although this result is uncertain due to quite significant low redshift contamination corrections, such features have been seen in lower redshift galaxy samples, particularly for LRGs at $z\approx0.5$ \citep[e.g.][]{ross07,2011MNRAS.416.3033S}. Given the improved power of the VLRS, it is interesting to see if there is any evidence of a change in the slope at small scales in $\xi(s)$ and $w_p(\sigma)$ in our $z\approx3$ LBG sample.

We therefore show in Fig.~\ref{fig:wpsig} a double power-law model for
$w_p$  with the same power-law slopes as fitted by
\citetalias{2011MNRAS.414....2B} to the LBG $w(\theta)$. We have reduced
the amplitude by $\approx20\%$ to match approximately the large-scale
amplitude fitted to the VLRS and Keck combined data. This is within the
systematic  uncertainties of the $w(\theta)$ measurement. Although
certainly not required  by the $w_p$ data this double power-law cannot
be rejected by the combined $w_p$ data, giving a reduced $\chi^2$ of
1.77 (marginally smaller than the reduced $\chi^2$ obtained for a single
power law of 1.84). 

In Fig.~\ref{fig:xis} we now compare to $\xi(s)$ the same double power-law $w(\theta)$ model with the $\approx20\%$ reduced amplitude. Again with a velocity dispersion of $\vdispar\kps$ and $\beta=\infpar$ we see that the model cannot be rejected by the data. We note that if we use a $\left<DD\right>/\left<DR\right>$ estimator the VLRS $\xi(s)$ result shows increased power at  large scales and the flatter slope of the double power-law model here provides a better fit.

We note that other authors have also reported a turn-up in the clustering at small scales in high redshift galaxy samples. For instance \citet{2005ApJ...635L.117O} reports that $z=4$ LBG $w(\theta)$ shows a steepening below $\approx0.2\hmpc$ or $\approx 10''$ at $z=4$. If both results are unaffected by contamination  then it could argue for an evolutionary growth in this break scale between $z=4$ and $z=3$.

Certainly there is plenty to motivate expanding surveys to make more
accurate measurements of both the angular and redshift survey
correlation functions at these redshifts. Below the break scale is of
extreme interest for single halo galaxy formation models and at large
scales the interest is in looking for a flattening of the correlation
function slope due to the presence of primordial non-Gaussianity.

\subsection{Estimating $\Omega_m$ and the growth rate}

\subsubsection{The mass density of the Universe}

We now look at the cosmological results afforded by the $z\approx3$ LBG
clustering and dynamics. As discussed by \citet{2002MNRAS.332..311H,daangela05a}, it is, in
principle, possible to constrain the matter density $\Omega_m(z=0)$ from
the measurement of $\xi(\sigma,\pi)$. Effectively, the elongation of
$\xi(\sigma,\pi)$ along the line of sight increases with increasing
values of $\Omega_m(z=0)$. However, increased values of $\beta$ lead to
a flattening of $\xi(\sigma,\pi)$ along the line of sight. These
effects combined lead to a degeneracy in determining $\Omega_m$ from the
galaxy clustering alone.

\begin{figure}
\centering
\includegraphics[width=8.0cm]{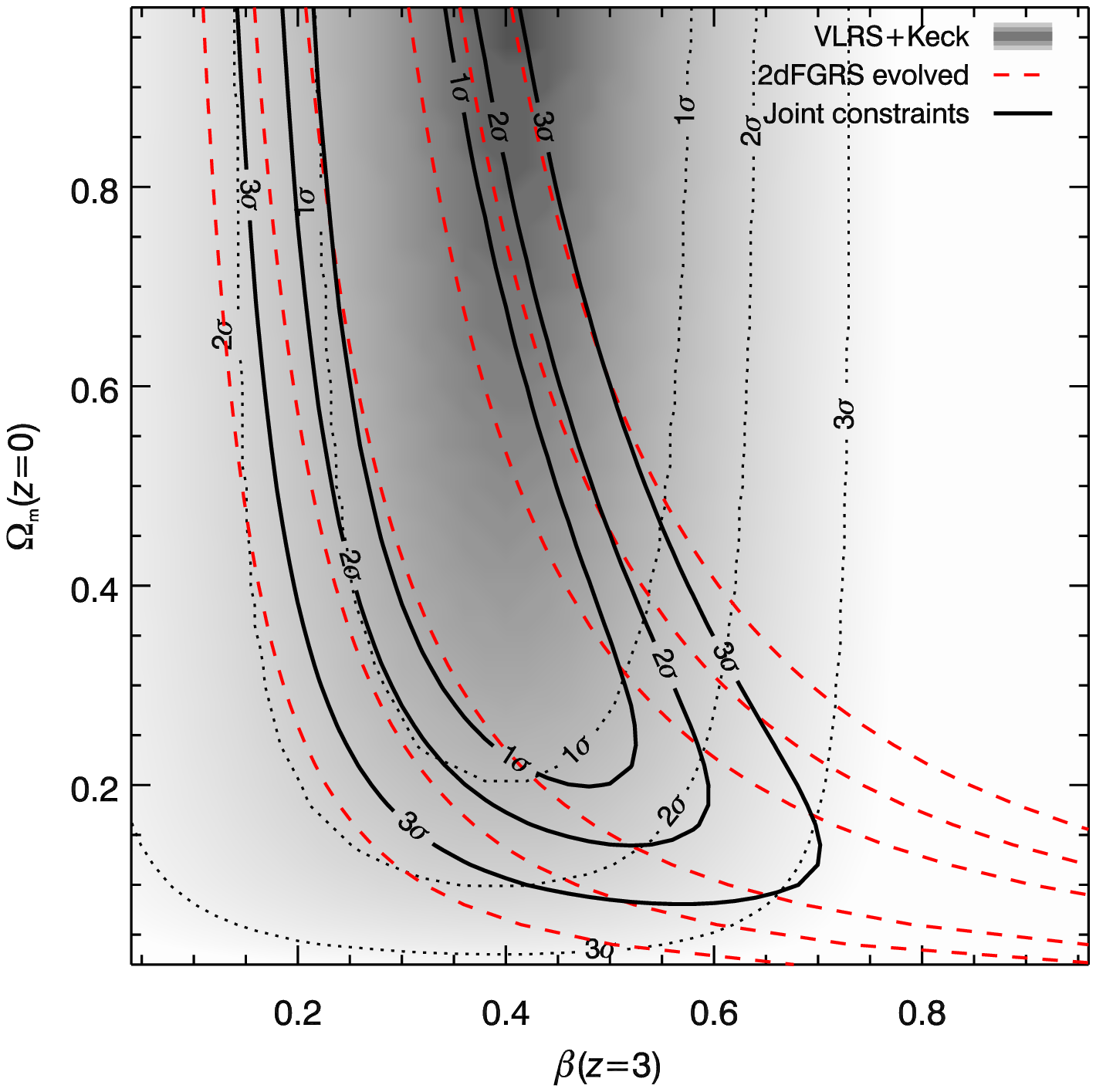}
\caption[Redshift measurement errors]{Fitting contours for the mass
density and bulk inflow based on the combined VLRS$+$Keck
$\xi(\sigma,\pi)$. The shaded region gives the result from the
VLT$+$Keck data sample (assuming $\vdisp=\vdispar\kps$,
$r_0=\cwprn\hmpc$ and $\gamma=\cwpgam$) with the dotted contour lines
giving the 1, 2 and $3\sigma$ uncertainties. The results from this fit
are $\beta(z=3)=\infparoma^{\infparomaup}_{\infparomadown}$ and
$\Omega_m(z=0)=\ompara^{\omparaup}_{\omparadown}$. The dashed red lines
show the 1, 2 and $3\sigma$ constraints given by evolving the 2dFGRS
measurements as described in the text. The solid black contours give the
combination of the two and give a result of
$\beta(z=3)=\infparomb^{\infparombup}_{\infparombdown}$ and
$\Omega_m(z=0)=\omparb^{\omparbup}_{\omparbdown}$.}
\label{fig:om_beta_fit}
\end{figure}

In previous sections, we have studied  the galaxy dynamics assuming a
cosmology with $\Omega_m(z=0)=0.3$. We now fit the $\xi(\sigma,\pi)$
result with this assumed cosmology, but now with a constant peculiar
velocity of $\vdisp=\vdispar\kps$ and fitting for $\Omega_m(z=0)$ and
$\beta$. The result is shown by the $1\sigma$, $2\sigma$ and $3\sigma$
contours (solid)  in Fig.~\ref{fig:om_beta_fit}. Based on just the $z=3$
galaxy clustering, we find results for  the mass density of
$\Omega_m(z=0)=\ompara^{\omparaup}_{\omparadown}$ and on the infall
parameter of $\beta(z=3)=\infparoma^{\infparomaup}_{\infparomadown}$.

Breaking the degeneracy of this result can be achieved by incorporating lower redshift results as shown by \citet{daangela05b,daangela05a} and \citetalias{2011MNRAS.414....2B}. As in these previous works, we use the 2dFGRS measurements of \citet{hawkins03} to do this ($r_0=5.0\hmpc$, $\gamma=1.8$ and $\beta(z=0.11)=0.49\pm0.09$). The \citet{hawkins03} result can then be evolved to the redshift of our study based on the relationship between the growth parameter, $f(z)$, and the bulk motion and the clustering bias, $b$, of a galaxy population:

\begin{equation}
\beta = \frac{f(z)}{b} \approx \frac{\Omega_m(z)^{0.55}}{b}
\label{eq:fz}
\end{equation}

The bias can be calculated directly from the clustering measurements by using the volume averaged clustering:

\begin{equation}
b=\sqrt{\frac{\xi_g(s)}{\xi_{DM}(s)}}=\sqrt{\frac{\overset{\_}{\xi}_g(8)}{\overset{\_}{\xi}_{DM}(8)}}
\label{eq:bias}
\end{equation}

\noindent where $\overset{\_}{\xi}_g(8)$ is the volume averaged correlation function at $s<8 \hmpc$ for the galaxy population and $\overset{\_}{\xi}_{DM}(8)$ is the same, but for the underlying dark matter distribution. The volume averaged clustering is calculated from the clustering using:

\begin{equation}
\overset{\_}{\xi}(x) = \frac{3}{x^3}\int^{x}_{0}r^2\xi(r)\mbox{d}r
\end{equation}

In addition, a measure of the dark matter clustering is required in order to estimate the bias of the galaxy population and we calculate this using the {\small CAMB} software incorporating the {\small HALOFIT} model of non-linearities \citep{smith03}. Using the previously determined best fit parameters of $r_0=\cwprn\hmpc$ and $\gamma=\cwpgam$, we evaluate the galaxy bias based on a single power-law, finding a bias for the LBGs of $b=\cbias\pm\cbiase$.

We then determine the $z=0.11$ underlying dark matter clustering amplitude from these parameter constraints and evolve this to $z=3$ for test cosmology range of $\Omega_m(z=0)=0-1$. The constraints on $\beta$ using this method over a range of assumed $\Omega_m$ values are given by the red dashed contours in Fig~\ref{fig:om_beta_fit}. By combining these with the original constraints from $\xi(\sigma,\pi)$, we find a result of $\beta(z=3)=\infparomb^{\infparombup}_{\infparombdown}$ and $\Omega_m(z=0)=\omparb^{\omparbup}_{\omparbdown}$.

Across these analyses, we have consistently found a value for the infall parameter of $\beta(z=3)\approx0.36-0.40$. $\Omega_m$ is somewhat less well constrained, but remains consistent with $\Lambda$CDM. The measurements of $\beta(z=3)$ presented here are consistent with our previous measurement from \citetalias{2011MNRAS.414....2B} of $\beta(z=3)=0.48\pm0.17$, whilst being somewhat higher than the result found by \citet{daangela05b} of $\beta(z=3)=0.15^{+0.20}_{-0.15}$. We note that the latter assumes a fixed velocity dispersion of $\vdisp=400\kps$ and is limited to the small field of view of the Keck survey As such, their lower estimate of $\beta$ may well be a systematic of too small an area to identify the Kaiser effect as well as not being able to simultaneously fit for the velocity dispersion.

\subsubsection{Growth rate results compared}

Using the results for $\beta$ and the galaxy bias we can compare our constraints of the growth parameter $f(z)$ to previous results. \citet{2008Natur.451..541G} presented the results of such an analysis based on the VLT VIMOS Deep Survey (VVDS), showing values for $f(z)$ extracted from a number of galaxy surveys up to a redshift of $z\approx0.8$. Here we add the $z\approx3$ result from our survey. We present measurements in terms of both $f(z)$ and $f\sigma_8$, where $f\sigma_8$ is intended to give a measurement which is \emph{less} dependent on the cosmology assumed for the calculation of the clustering \citep[e.g.][]{2009JCAP...10..004S}.

\begin{figure}
\centering
\includegraphics[width=8.0cm]{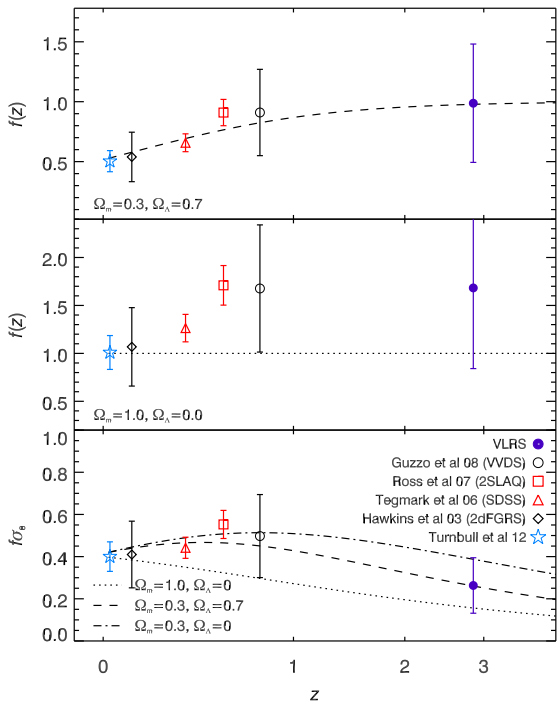}
\caption[Growth factor]{The evolution of the growth factor based on
available $z<1$ survey observations and the constraint from the VLRS
data (solid blue circle). }
\label{fig:growth}
\end{figure}

We have already calculated the infall parameter and take the value ($\beta=\infpar$) obtained via fitting the velocity dispersion and $\beta$ in a $\Lambda$CDM cosmology with $\Omega_m=0.3$ and
$\Omega_\Lambda=0.7$ (Fig.~\ref{fig:vdisp_beta_fit}). Combining this with our measurement of the galaxy bias gives a value for the growth parameter based on the combined LBG sample of $f(z=3)=\fpar\pm\fpare$.

We present the $f(z=3)$ result (filled blue circle) in the top panel of Fig.~\ref{fig:growth} alongside a number of other low-redshift measurements. In order of ascending redshift, the star shows the measurement of \citet{2012MNRAS.420..447T} based on local supernovae measurements, the diamond shows the result based on the 2dFGRS presented by \citet{hawkins03}, the red triangle shows the SDSS result based on LRGs from \citet{2006PhRvD..74l3507T}, the red square shows the 2SLAQ result also estimated from the LRG population of \citet{ross07} and the black circle shows the VVDS result of \citet{2008Natur.451..541G}. For completeness these are also summarised in Table~\ref{tab:growthpara}.

\begin{table*}
\caption{Summary of growth parameter results from the literature.}
\begin{threeparttable}[b] 
\label{tab:growthpara}
\begin{tabular}{lccccc}
\hline
Survey                              & $z$   & $b$           & $\beta$       & $f$           & $f\sigma_8$ \\
\hline
First Amendment SNe\footnotemark[1] & 0.025 & ---           & ----          & ---           & $0.40\pm0.07$ \\
2dFGRS\tnote{2}              & 0.11  & $1.15\pm0.06$ & $0.47\pm0.18$ & $0.54\pm0.21$ & $0.41\pm0.16$ \\
SDSS LRGs\footnotemark[3]           & 0.35  & $2.13$\footnotemark[4] & $0.31\pm0.04$ & $0.66\pm0.07$ & $0.44\pm0.05$ \\
2SLAQ LRGs\footnotemark[5]          & 0.55  & $2.02\pm0.10$ & $0.45\pm0.05$ & $0.91\pm0.11$ & $0.55\pm0.07$ \\
VVDS\footnotemark[6]                & 0.77  & $1.30\pm0.10$ & $0.70\pm0.26$ & $0.91\pm0.07$ & $0.50\pm0.04$ \\
VLRS$+$Keck                     & 2.85  & $\cbias\pm\cbiase$ & $\infpar\pm\infpare$ & $\fpar\pm\fpare$ & $\fsig\pm\fsige$\\
\hline
\end{tabular}
\begin{tablenotes}[para]\footnotesize 	
\item[1]\citet{2012MNRAS.420..447T}
\item[2]\citet{hawkins03}
\item[3]\citet{2006PhRvD..74l3507T}
\item[4]\citet{2009JCAP...10..004S}
\item[5]\citet{ross07}
\item[6]\citet{2008Natur.451..541G}
\end{tablenotes}
\end{threeparttable}
\end{table*} 

In the middle panel of Fig.~\ref{fig:growth}, we also plot the evolution
of $f(z)$ based on the assumed $\Lambda$CDM cosmology (dashed line),
where $f(z)=\Omega_m(z)^{0.55}$. The low redshift data points are all
consistent with the assumed cosmology at the $\sim1\sigma$ level and at
$z=3$, the model cosmology is again consistent with the data. We note
again that the observations themselves depend on the assumed cosmology
via $\sigma_8(z)$ and so to test the $\Omega_m=1$ cosmology we adjust
the observed values of $f(z)$ for the effects of different cosmology in
eq. \ref{eq:bias} according to the methods set out by 
\citet{daangela05b}. Also assuming that $\beta$ is approximately
independent of the assumed cosmology, we see that the $\Omega_m=1$
$z$-independent growth rate  is apparently rejected by the data.
However, if the bias is allowed to float rather than just fit the lowest
redshift point then the model may only be rejected at the $1-2\sigma$
level, consistent with the conclusions from Fig. \ref{fig:om_beta_fit}.

If we now consider $f\sigma_8$, the observations are now
independent of the assumed cosmology, at least given again the
assumption that the observed $\beta$ is approximately cosmology
indpendent. Each of the observational measurements is again plotted in
the top panel of Fig.~\ref{fig:growth}, but now in terms of $f\sigma_8$.
We now plot three test cosmologies for comparison, the $\Lambda$CDM used
in the top panel (dashed line), plus an Einstein-de-Sitter model
($\Omega_m=1$, $\Omega_\Lambda=0$, dotted line) and an open Universe
without a cosmological constant and a mass density of $\Omega_m=0.3$
(dot-dash line). For each model we incorporate a factor ($c_{at}$) to 
correct for the cosmology assumed in the measurement of the clustering
observations being different from the test cosmology. Each model is thus
given by:

\begin{equation}
(f\sigma_8)_a = \beta\sigma_{g,a} = \frac{\beta\sigma_{g,t}}{\sqrt{c_{at}}} = \frac{(f\sigma_8)_t}{\sqrt{c_{at}}} = \frac{\Omega_{m,t}(z)^{0.55}\sigma_{8,t}(z)}{\sqrt{c_{at}(z)}}
\end{equation}

\noindent where an index of $t$ denotes a parameter calculated in the
test cosmology and an index of $a$ denotes a parameter calculated in the
assumed cosmology (i.e. $\Lambda$CDM). We normalise $\sigma_8$ to 0.8 at
$z=0$ and $\sigma_g$ is effectively $\sigma_8$ measured for the galaxy
population. If we assume a power-law form for the clustering with a
slope of $\gamma=1.8$, then following
\citet{1996MNRAS.282..877B,daangela05b} $c_{at}$ is given by:

\begin{equation}
c_{at} = \left(\left(\frac{B_t}{B_a}\right)^2\frac{A_t}{A_a}\right)^{2/3}
\end{equation}

\noindent with $A$ and $B$ \citep{daangela05b} given by:

\begin{equation}
A = \frac{c}{H_0}\frac{1}{\Omega_\Lambda^0+\Omega_m^0(1+z)^3}
\end{equation}

\begin{equation}
B = \frac{c}{H_0}\int_0^z\frac{\mbox{d}z'}{\Omega_\Lambda^0+\Omega_m^0(1+z')^3}
\end{equation}

We are assuming here that $\beta$ is independent of cosmology, a reasonable 
approximation  when $\Omega_m\ga0.1$

With the models corrected to account for differences between
the cosmology assumed for the observations and the model cosmologies,
the observations now provide clearer tests on the models. We note
that the observations (excepting the SDSS point) assume the
$\Omega_m=0.3$/$\Omega_\Lambda=0.7$ cosmology and so there is no change
in the relationship between the observations and the $\Lambda$CDM model
between the top and bottom panels. Thus the VLRS datapoint shows the
same level of consistency with the $\Lambda$CDM model for $f\sigma_8$
and $f(z)$.

At redshifts of $z<1$, we see that flat and open cosmologies (i.e. with
and without a cosmological constant) are poorly distinguished by the
available observations. At $z=3$, we find that the VLRS data can only reject the
open cosmology with $\Omega_m=0.3$ at the $\approx1\sigma$ level. The
Einstein-de Sitter cosmology is  apparently rejected by combining the
$z<1$ and $z\approx3$ observations. But again if the normalisation of the
model is allowed to float rather than be fixed on the low-redshift
SNe observation, the model still fits the data with a reduced $\chi^2$ 
of 2.7.

\subsection{The dark matter halos of $z\sim3$ star-forming galaxies}

We now look at the nature of the halos that host the LBG sample based on our clustering results using the halo-occupation distribution (HOD) model formalism \citep[e.g.][]{2000ApJ...543..503M,2000MNRAS.318.1144P,2004ApJ...608...16Z,2005ApJ...633..791Z}. By matching our clustering results to the measured clustering properties of simulated dark matter halos, we estimate mean halo masses ($\left<M_h\right>$), minimum halo masses ($M_{min}$) and cumulative occupation numbers ($\left<N(M)\right>$) for the VLRS, Keck and combined samples. The simulation results are obtained from the cosmological dark matter simulation described by \citet{2010MNRAS.407.1449G} and the results are given in Table~\ref{tab:halos}.

\begin{table*}
\centering
\caption{Results of the halo matching analysis.}
\label{tab:halos}
\begin{tabular}{lcccc}
\\
\hline
Sample   & Bias               & log($\left<M_h\right>$/$h^{-1}$M$_\odot$) & log($M_{min}$/$h^{-1}$M$_\odot$) & $\left<N(M)\right>$ \\ 
\hline
VLRS     & $\vbias\pm\vbiase$ & $\vhmass\pm\vhmasse$ & $\vhminm\pm\vhminme$ & $\vhocc\pm\vhocce$  \\ 
Keck     & $\kbias\pm\kbiase$ & $\khmass\pm\khmasse$ & $\khminm\pm\khminme$ & $\khocc\pm\khocce$  \\ 
Combined & $\cbias\pm\cbiase$ & $\chmass\pm\chmasse$ & $\chminm\pm\chminme$ & $\chocc\pm\chocce$  \\ 
\hline 
\\
\end{tabular}
\end{table*}

We find that the results are consistent within the error bars between the VLRS and Keck samples evaluated separately, with mean host halo masses of $\sim10^11.6~h^{-1}$M$_\odot$. This lends additional credence to our having combined the two samples in order to improve the statistical fidelity of the 2D clustering results. In addition we note that the occupation numbers suggest that multiple LBGs are present in single galaxies, with $\left<N(M)\right>$ consistently $>1$ for all the samples although this is with relatively large uncertainties.

There are few other measurements of the halo masses of $z\sim3$ LBGs available in the literature that are based on spectroscopic data. In terms of photometric samples, \citet{foucaud03}, \citet{hildebrandt07} and \citet{2008ApJ...679..269Y} measure halo masses of bright $z\approx3$ LBG samples of $M_{DM}\sim10^{12}h^{-1}\rm{M_\odot}$, an order of magnitude larger than for our sample. However, in a similarly photometric study, \citet{2006ApJ...642...63L} found marginally lower halo masses of $\sim5-10\times10^{11}h^{-1}\rm{M_\odot}$ for both $z\sim3$ and $z\sim4$ LBGs, the $z\sim3$ LBGs having a magnitude limit of $r=25.5$. Similarly, the results of \citet{2012ApJ...752...39T} show a halo mass of $M_{DM}\sim10^{11.9\pm0.1}h^{-1}\rm{M_\odot}$, but is based on galaxies with a redshift distribution somewhat lower than our own.

In terms of the spectroscopic samples closest in redshift and form to our own, work using the \citet{steidel03,2004ApJ...604..534S} data report halo masses of $M_{DM}\sim10^{11.5\pm0.3}h^{-1}\rm{M_\odot}$ \citep{2005ApJ...619..697A}. These spectroscopic $z\sim3$ based measurements are in good agreement with our own results. This is as one would expect for our `Keck' sample given that this uses some of the same data as the above results, whilst the consistency between these results and the result from our own pure-VLRS sample adds weight to the results as a whole. As noted by previous authors, the LBG host halo masses are approximately an order of magnitude lower than those measured for the infrared selected population at $z\sim3$ \citep[e.g.][]{2007ApJ...654..138Q}, hinting at the continued trend for a `blue' star-forming population existing in low-density environments and a `red', potentially more passive population inhabiting denser environments.

\section{Conclusions}

We have presented the widest area spectroscopic survey of galaxies thus far in the redshift range $2<z<3.5$, based on observations with the VLT VIMOS instrument. This paper adds to the initial dataset of \citet{2011MNRAS.414....2B}, where data in five 0.5\deg\ $\times$ 0.5\deg\ fields were presented. Here we add a further four new fields, each with deep optical imaging over an area of $\approx$ 0.5\deg $\times$ 0.5\deg\ in three cases and a full 1\deg\ $\times$ 1\deg\ in the fourth field. In addition, we have extended one of the original fields of \citetalias{2011MNRAS.414....2B} to 1\deg$\times$ 1\deg\ from the original 0.5\deg\ $\times$ 0.5\deg. In total therefore, we now have $\approx4$ deg$^2$ of optical imaging with a minimum of three bands in each field incorporating  $U$, $B$ and $R$ or equivalents.

In total, the survey now consists of \numlbg\ spectroscopically confirmed $z>2$ galaxies. The properties of the full sample have been presented here with redshift and magnitude distributions as well as example and composite spectra. The mean redshift of our $z>2$ galaxy dataset is $\bar{z}=\meanz$. In addition, we detect 30 AGN or quasars, $\approx800$ low-redshift galaxies and $\approx$130 Galactic stars. Using the $z>2$ galaxy dataset, we have conducted an analysis of the galaxy clustering at $z\sim3$. Using the semi-projected correlation function, we have measured a galaxy clustering length of $r_0=\vwprn\pm\vwprne\hmpc$ with a slope of $\gamma=\vwpgam\pm0.26$, assuming a power-law form to $\xi(r)$. We have also combined the VLRS sample with the Keck LBG sample of \citet{steidel03}, which provides greater statistical power on small
scales (i.e. $s\lesssim2\hmpc$) than the VLRS but does not provide the
coverage of the VLRS at larger scales (i.e. $s\gtrsim8\hmpc$). For the
combined sample we measure a clustering length of $r_0=\cwprn\pm\cwprne\hmpc$, with a slope of $\gamma=\cwpgam\pm\cwpgame$.

We have shown that the LBG correlation functions consistently show slopes that are significantly flatter than the canonical $\gamma=1.8$ observed at low redshift. Indeed, the measured slopes of $\gamma=1.55$ are flatter than in some $z\approx1$ galaxy and radio-source correlation functions that have been interpreted as showing evidence for primordial non-Gaussianity \citep{2010ApJ...717L..17X, 2011PhRvL.106x1301T,2011MNRAS.416.3033S, 2012arXiv1204.3609N}. Non-Gaussianity is expected to be easier to detect at large-scales and high redshift. We have therefore checked whether a standard $\Lambda$CDM model is consistent with the form of the VLRS $\xi(s)$ in particular in the regime $10<s<50\hmpc$. We found that there is evidence that the LBGs are showing more large-scale power than the standard model in this regime but only at $\approx2\sigma$. More studies of LBG clustering at large-scales are clearly needed to check these results.

In addition to the 1D clustering analyses, we have also investigated the 2D correlation function and the imprints of galaxy dynamics on the clustering. We find that the 2D clustering for the VLRS$+$Keck LBG sample is well fit by a model based on a power-law fit with a clustering length of $r_0=\vwprn\hmpc$, a large scale infall parameter of $\beta=\infpar\pm\infpare$ and a velocity dispersion of $\vdisp=\vdispar^{\vdisparup}_{\vdispardown}\kps$, over a range of $1<s<25\hmpc$. We have shown that this result is consistent with the model for the redshift space correlation function, $\xi(s)$, measured for the combined sample.

We use the 2D galaxy clustering results to determine the matter density
parameter and the growth parameter. Using the previously constrained form for the clustering and galaxy velocity dispersion, we fit the 2D correlation function for the matter density, $\Omega_m$. We find an acceptable range in the matter density of $\Omega_m(z=0)=\ompara^{\omparaup}_{\omparadown}$ (with an infall parameter of $\beta(z=3)=\infparoma^{\infparomaup}_{\infparomadown}$).
We add a further constraint provided by the 2dFGRS low-redshift
clustering measurements, which gives
$\Omega_m(z=0)=\omparb^{\omparbup}_{\omparbdown}$ (with an infall
parameter of $\beta(z=3)=\infparomb^{\infparombup}_{\infparombdown}$).
Although the constraints on the mass density are relatively weak, we see
that the constraints on the infall parameter remain consistent. Using
these measurements to constrain the growth parameter, we find a value of
$f(z=3)=b\beta=\fpar\pm\fpare$. In addition we determine the combined
parameter $f\sigma_8$, which gives a measure of the growth parameter
that is less dependent on the assumed underlying dark matter mass
distribution. In this case we find a value of $f\sigma_8=\fsig\pm\fsige$.
These measurements are the highest redshift constraint on the growth
parameter based on galaxy clustering analyses. We have shown that these
measurements are consistent with the $\Lambda$CDM standard model,
although given the uncertainties on the measurements, they are also
consistent with a number of other cosmologies.

Based on the clustering results, we estimate typical halo masses for the dark matter halos that host the LBG population. For the VLRS sample alone, we estimate a mean halo mass of $M_{DM}=10^{\vhmass\pm\vhmasse}$, consistent with measurements based on comparable spectroscopic LBG samples at $z\sim3$ and an order of magnitude lower than the typical halo masses hosting $z\sim3$ infra-red selected galaxies.

This work is one of the largest surveys of the galaxy mass distribution
at $z\approx3$ and paves the way for a number of lines of research,
which will be followed in subsequent papers. In particular, the
proximity of the data presented here to quasar sightlines will provide
important constraints on the relationship between galaxies and the IGM
at an epoch associated with significant interactions between the two.

\section*{Acknowledgments}

We thank Mike Irwin for assistance with the WFCAM data reduction. This
work was based on data obtained with the NOAO Mayall 4m Telescope at
Kitt Peak National Observatory, USA (programme ID: 06A-0133), the NOAO
Blanco 4m Telescope at Cerro Tololo Inter-American Observatory, Chile
(programme IDs: 03B-0162, 04B-0022) and the ESO VLT, Paranal, Chile
(programme IDs: 075.A-0683, 077.A-0612, 079.A-0442). MDH acknowledges
the support of a STFC PhD Studentship grant, whilst RMB, TS and NM also
acknowledge STFC funding. RMB acknowledges support from a grant obtained
from the Agence Nationale de la Recherche (ANR, France). This work was
partially supported by the Consejo Nacional de Investigaciones
Cient\'ificas y T\'cnicas and Secretaria de Ciencia y T\'cnica de la
Universidad Nacional de C\'ordoba, and the European Union Alfa II
Programme, through LENAC, the Latin American-European Network for
Astrophysics and Cosmology. DM and LI are supported by FONDAP CFA
15010003, and BASAL CATA PFB-06. This research has made use of the
NASA/IPAC Extragalactic Database (NED) which is operated by the Jet
Propulsion Laboratory, California Institute of Technology, under
contract with NASA.

\bibliographystyle{mnras_mod}
\bibliography{$HOME/Documents/lib/rmb}

\label{lastpage}

\end{document}